\documentclass{aa}  
\newcommand{\rev}[1]{{#1}}
\newcommand{\revv}[1]{{#1}} 

\usepackage{graphicx}
\usepackage{xcolor}
\usepackage{txfonts}
\usepackage[hidelinks]{hyperref}
\usepackage{multirow}   
\usepackage{lscape,wasysym}

\begin{document}

   \title{Supersoft, luminous X-ray sources in galactic nuclei}

   \subtitle{}

   \author{A. Sacchi
          \inst{1,2,3}
          \and
          G. Risaliti\inst{1,4}
          \and
          G. Miniutti\inst{5}
          }

   \institute{Dipartimento di Fisica e Astronomia, Università di Firenze, via G. Sansone 1, I-50019 Sesto Fiorentino, Firenze, Italy\\
              \email{andrea.sacchi@iusspavia.it}
              \and
              Istituto universitario di Studi superiori di Pavia, Palazzo del Broletto, Piazza della Vittoria 15, I-27100 Pavia, Italy
              \and
              Instituto de F\'isica de Cantabria (CSIC-UC), Avenida de los Castros, E-39005 Santander, Spain
              \and
              INAF – Osservatorio Astrofisico di Arcetri, Largo Enrico Fermi 5, I-50125 Firenze, Italy
              \and
              Centro de Astrobiología (CAB), CSIC-INTA, Camino Bajo del Castillo s/n, E-28692, Villanueva de la Ca\~n ada, Madrid, Spain
              }         

   \date{Received September 15, 1996; accepted March 16, 1997}

 
  \abstract
{Tidal disruption events (TDEs) are usually discovered at X-ray or optical wavelengths through their transient nature. A characteristic spectral feature of X-ray detected TDEs is a "supersoft" X-ray emission, not observed in any other extragalactic source, with the exception of a few, rapidly variable hyper-luminous X-ray sources (HLXs) or supersoft active galactic nuclei (AGN) that are however distinguishable by their optical emission.}
   {The goal of our work is to find extragalactic supersoft sources associated with galactic centres. We expect this category to include overlooked TDEs, supersoft AGN and nuclear HLXs. Finding such sources would allow for the study of extreme regime accretion on different black hole mass scales. }
   {\revv{We searched for supersoft X-ray sources (SSS) by cross-correlating optical and X-ray catalogues to select extragalactic near-nuclear sources and we then filtered for very steep spectra (photon index $\Gamma>3$) and high X-ray luminosities ($L_X>10^{41}$~erg~s$^{-1}$).}}
   {With our blind search we retrieved about 60 sources including 15 previously known supersoft AGN or TDEs, so demonstrating the efficiency of our selection. Of the remaining sample, 36 sources, although showing steeper-than-usual spectra, are optically classified as AGN. The remaining nine, previously unknown sources show spectral properties consistent with the emission by extremely soft-excess dominated AGN (five sources) or TDE (four sources). An {\it XMM-Newton} follow-up observation of one of these sources confirmed its likely TDE nature.}
   {Our work is the first attempt to discover TDEs by their spectral features rather than their variability and has been successful in retrieving known TDEs as well as discovering new extreme ultrasoft sources, including four new TDE candidates, one of which is confirmed via follow-up observations.}


   \maketitle
%

\section{Introduction}

Supersoft X-ray sources (SSSs) are objects emitting in the X-ray band with typical effective temperatures  kT$\sim20-100$~eV and observed luminosities rarely exceeding a few times $10^{38}$~erg~s$^{-1}$, but well above what can be produced in stellar coronae \rev{ \citep{mcl42,mcl43}}. Traditionally, SSSs are mostly associated with accreting white dwarfs (WD) where mass transfer leads to nuclear fusion on the WD surface either in the form of runaway events (as in classical novae) or in a steady manner for some systems with high mass accretion rate. Today $\sim80$ novae have been observed in the X-ray band including extragalactic ones in the Magellanic Clouds, M31 and M33 (see \citealt{del20} and references therein for a review). 

However, novae and/or steady nuclear burning WD are not the only known extragalactic SSS, and some intriguing classes of X-ray-emitting objects appear to sometimes fall in this very same category. In particular, possible high-luminosity members of this class may include soft ultraluminous X-ray sources (ULX) \rev{\citep{swa04,liu05,wal11,kaa17}}, tidal disruption events (TDEs) \rev{\citep{kom15,auc18,sax20,gez21}}, and supersoft active galactic nuclei (AGN) whose X-ray luminosity is almost entirely emitted below $\sim$~2~keV \rev{\citep{ter12,sun13}}. Heavily obscured AGN may also predominantly shine in the soft X-ray band because reprocessed/scattered emission on large scales can easily dominate over the heavily absorbed intrinsic emission for large enough column densities, \rev{although the X-ray emission is usually not as steep as in the other classes of objects \citep{gua99,com04,bia06,gua07}}. 

Ultraluminous X-ray sources (ULXs) are defined as off-nuclear X-ray sources with luminosities exceeding $10^{39}$~erg~s$^{-1}$. Detailed observations and modelling in recent years favour an interpretation of the ULX phenomenon in terms of supercritical accretion onto stellar mass black holes and neutron stars (e.g. \citealt{mid12,mid15,bac14}). Unlike typical ULXs which are often associated with  significant emission above 1-2~keV, a sub-class of sources dubbed ultraluminous supersoft X-ray sources (ULSs) are dominated by a thermal-like component with a typical effective temperature of $50-200$~eV possibly due to reprocessing in an optically thick outflow blocking most/all hard X-ray photons from view (see e.g. \citealt{urq16,fen16,pin17}). On the other hand, the most luminous examples of ULXs ($L_{\rm{X}}\gtrsim 10^{41}$~erg~s$^{-1}$), known as hyperluminous X-ray sources (HLXs), are of particular interest as they can be considered bona fide intermediate-mass black hole (IMBH) candidates (e.g. \citealt{far09}). As the thermal X-ray emission from accretion discs around IMBHs is naturally expected to dominate the soft X-ray band, accreting IMBH candidates are likely to be observed as hyperluminous ULSs. Most ULX (and HLX) searches so far are biased in favour of off-nuclear sources to avoid potential confusion with AGN, see e.g. \citet{bar19}. However, IMBHs are expected to be found also in the nuclear regions \revv{(e.g. \citealt{chi18})}.

TDEs occur when a star wanders too close to a massive black hole. The tidal forces of the hole tear the star apart and destroy it. The stellar debris gets subsequently partly ejected and partly accreted onto the hole generating a bright transient. The study of TDEs is of particular interest as this kind of event offers a unique way to investigate otherwise quiescent (non-AGN) supermassive black holes (SMBHs) that lurk in the centres of most known galaxies. The rapid evolution of the electromagnetic emission from TDEs \citep{ree88,phi89} also offers a unique opportunity to study different SMBH accretion regimes in the same source on human timescales. The optically thick accretion disc that the stellar debris should form around the hole is predicted to emit thermal radiation with a characteristic temperature of some tens of eV; this black body radiation would have a visible tail in the X-ray band, appearing as an extremely steep power-law spectrum. Indeed the first candidates of TDEs were found in the {\it ROSAT} All Sky Survey as bright transients with very soft X-ray spectra ($\Gamma\approx3-5$) \citep{bad96}. \rev{X-ray TDEs have typical peak X-ray luminosities in the range of ${\rm few} \times 10^{42}-{\rm few} \times 10^{44}$~erg~s$^{-1}$. Their high-luminosity X-ray spectrum is generally well described by thermal
spectral models with $kT = 40-80$~eV, often evolving into a harder spectral shape possibly due to an emerging hard X-ray power-law component that may signal the formation, as the Eddington ratio decays, of the optically-thin X-ray corona that characterizes the X-ray emission in AGNs above 1-2~keV \citep{wev20}. However, the spectral evolution is diverse, and some events actually evolve softening and or maintaining a roughly constant spectral shape (see \citealt{sax20} for a review).}  Today's most efficient methods employed to identify TDEs involve the detection of bright transients followed by multi-wavelength observations focused on determining their nature. The bright transients are usually spotted serendipitously, be it in the optical or X-ray band, by the comparison between old catalogues and new observations of large sky portions. Although these methods are efficient in finding new events, it is likely that several TDEs are hidden in available catalogues and archives, being spotted by chance at, or near, their luminosity peak.

\rev{In the X-ray band, the supersoft X-ray emission from TDEs is remarkably different from typical unobscured AGNs and can be used to distinguish between the two classes, at least at the zeroth-order. Indeed, AGNs have X-ray spectra that are typically well described with a power-law component with $\Gamma\sim1.9$ dominating above $1-2$ keV. In the soft X-rays, the spectrum is often dominated by a softer X-ray component that can be phenomenologically described with blackbody emission with typical $kT\sim 100-200$ eV \citep{pic05}. Reflection and absorption components are often superimposed on the general continuum shape described above. While the hard power law is most likely due to inverse Compton scattering  of the softened UV/EUV accretion disc photons in an optically-thin and hot ($kT\gtrsim100$ keV or so) X-ray corona, the origin of the soft excess is still debated, although models invoking Comptonization in an optically-thick warm ($kT\sim100-200$ eV) corona have recently attracted much attention \citep{don12,pet18}.}

Finally, a few X-ray supersoft AGNs with peculiar optical and X-ray properties have emerged in recent years. 2XMM~J123103.2+110648 \citep{ter12,ho12} is an optically classified Seyfert~2 galaxy with supersoft X-ray spectrum and significant X-ray variability, mostly associated with a claimed $\sim 3.8$~hr periodicity \citep{lin13}. Two other very similar objects, GSN 069 \citep{min13,min19} and RX J1301.9+2747 \citep{sun13,shu17,giu20}, exhibit spectacular recurrent X-ray variability in the form of short-lived and very high amplitude X-ray bursts occurring every few hours and superimposed to a quiescent constant level (quasi-periodic eruptions, or QPEs, see \citealt{min19,giu20}). In all three cases, optical spectra show Seyfert-like narrow emission lines, but no sign of broad lines which may indicate past activity and recent re-activation. QPEs have later been detected in two other sources with no sign of black hole activity in the optical by eROSITA, and then confirmed with {\it NICER} and {\it XMM-Newton} follow-up observations \citep{arc21}. Another QPE candidate was identified in a likely TDE source \citep{cha21}. All of these sources are associated with exceptional X-ray variability, no sign of X-ray absorption, supersoft X-ray spectra, and general properties pointing towards relatively low mass black holes ($10^5-10^6$~M$_\odot$) and high mass accretion rates. \rev{The peculiar behaviour both in the X-ray and optical band, i.e. the lack of broad optical emission lines in GSN~069 and RX~J1301.9+2747, may perhaps be understood if these sources were long-lived TDEs rather than AGN, and although this point is still being debated}, indeed there are convincing signs that this is the case at least for GSN~069  \citep{sun13,lin17a,shu18,she21}.

Here we present results from a search for SSS X-ray sources associated with the central region of external galaxies restricted to X-ray luminosities exceeding $10^{41}$~erg~s$^{-1}$.
Our search could in principle reveal a sample of new overlooked TDE, nearly nuclear supersoft HLXs, and a population of supersoft AGN.

\section{Source selection}

In order to identify interesting sources we started from the {\it XMM-Newton} catalogue of serendipitous sources (4XMM-DR9, \citealt{web20}). \rev{For each of the more than 8$\times10^5$ detections} we computed i) the signal-to-noise ratio in the ultrasoft and soft band (SNR$_{0.2-0.5~{\rm keV}}$ and SNR$_{0.5-1~{\rm keV}}$ respectively) and ii) the photon index for the ultrasoft to soft and soft to medium ($1-2$ keV) bands ($\Gamma_{0.2-1~{\rm keV}}$ and $\Gamma_{0.5-2~{\rm keV}}$ respectively). As the goal was to filter for reliable supersoft sources, we selected all detections with steep and high-quality spectra in the ultrasoft band i.e. $\Gamma_{0.2-1~{\rm keV}}$>3, SNR$_{0.2-0.5~{\rm keV}}$>7 and SNR$_{0.5-1~{\rm keV}}$>3. The selection in the ultrasoft band however allows for little to no absorption (even compared with the average Galactic one). To solve this issue we added to the previous selection all detections with steep and high-quality spectra in the soft band ($\Gamma_{0.5-2~{\rm keV}}$>3 and SNR$_{0.5-1~{\rm keV}}$>7), although, in order to avoid possible contamination by star-forming regions, dominated in the softer X-ray band by thermal/photo-ionized plasma emission, we also imposed a constraint on the ultrasoft to soft photon index: $\Gamma_{0.2-1~{\rm keV}}$>-1.

In order to obtain the redshift information we cross-matched our selected sources with four galaxy catalogues (SDSS DR16 \citep{ahu20}, 2dFGRS \citep{col03}, 6dFGS DR3 \citep{jon09} and LCRS \citep{she96}, listed by their dimensions), exploiting SIMBAD tools. With the distance information, we computed the X-ray luminosity for each detection in the full $0.2-12$ keV band and selected only those with $L_{\rm X}>10^{41}$ erg/s. This allows us to exclude possible Galactic SSS. The sample resulting from this selection is composed of 61 sources.

We furthermore excluded an additional source: SDSS J123408.85+090542.4, which is the brightest cluster galaxy (BCG) of SDSSCGA 1202, a compact cluster \citep{yoo08}. Spectral analysis of the soft X-ray emission of this galaxy revealed an optically thin, warm gas with solar abundances. This strongly suggests that its emission is due to gas warming up while falling in the potential well of the cluster, ruling out the possibility of accretion phenomena.

\begin{figure}
	\includegraphics[width=\hsize]{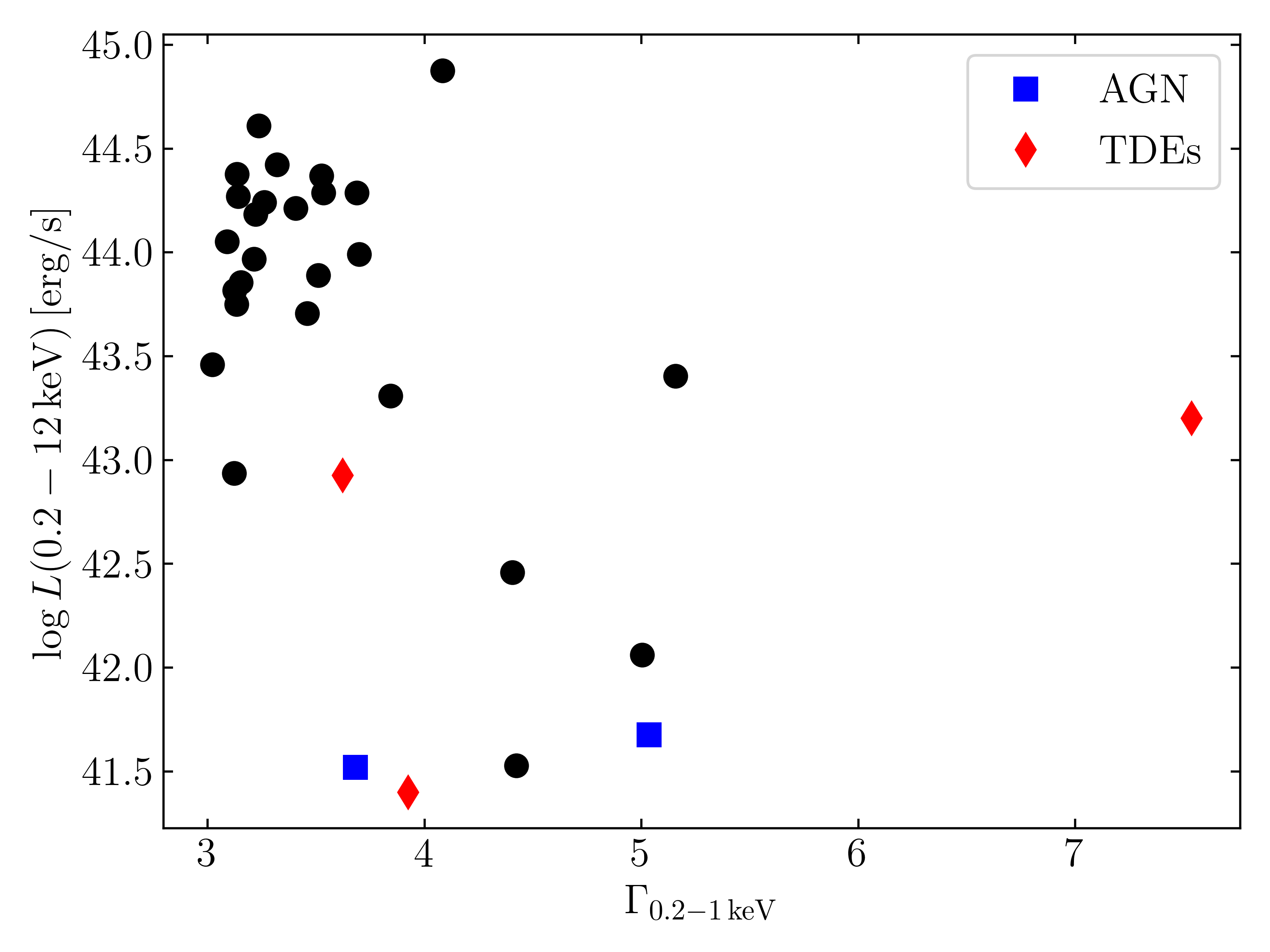}
    \caption{The 31 sources selected in the ultrasoft band shown in the $L_\textup{X}-\Gamma_{0.2-1~{\rm keV}}$ plane. Blue squares and red diamonds show the extreme AGN and TDEs known in the literature, respectively.}
    \label{fig:g21}
\end{figure}

\begin{figure}
	\includegraphics[width=\hsize]{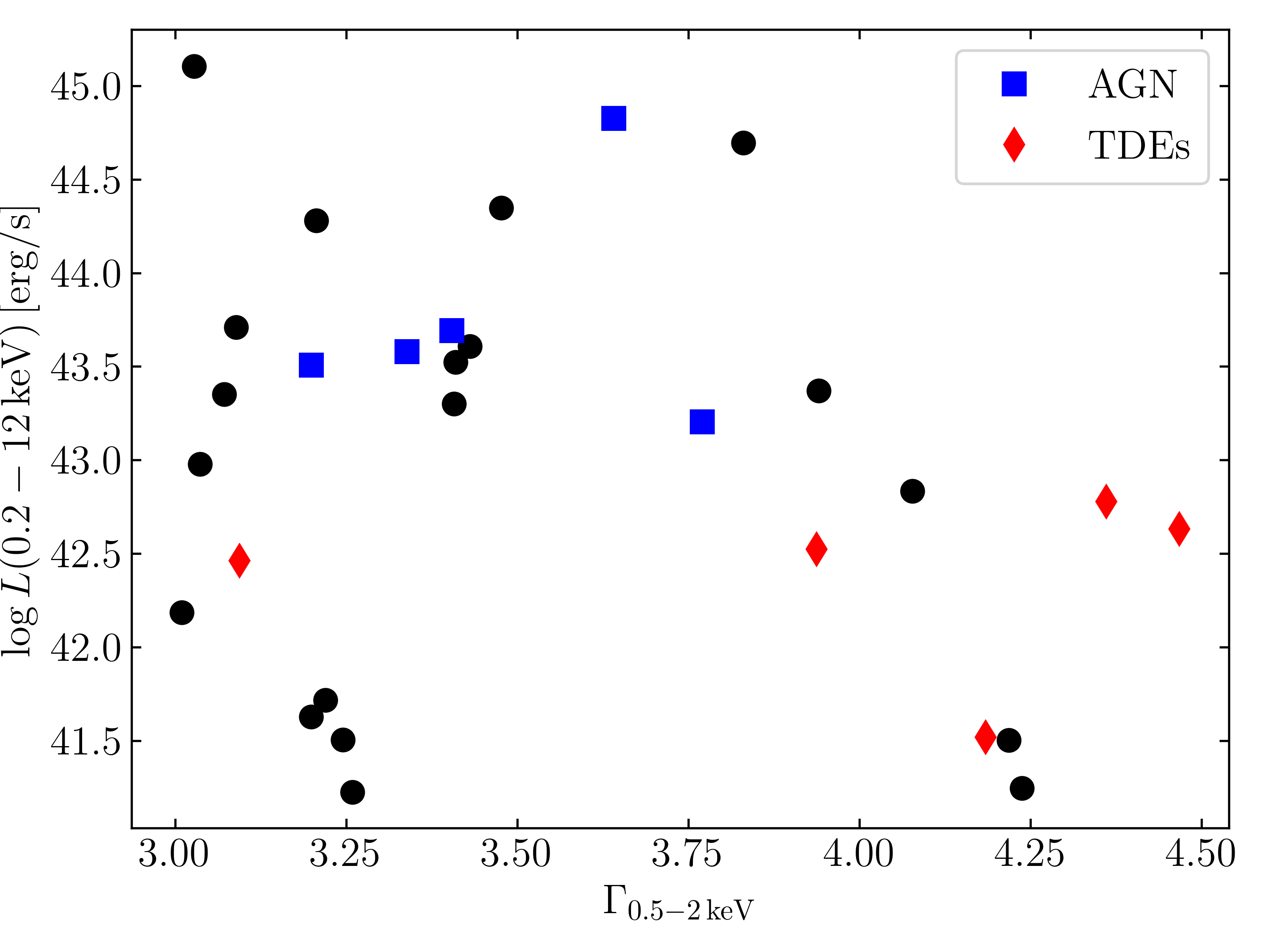}
    \caption{The 29 sources selected in the soft band shown in the $L_\textup{X}-\Gamma_{0.5-2~{\rm keV}}$ plane. Blue squares and red diamonds show the extreme AGN and TDEs known in the literature, respectively.}
    \label{fig:g32}
\end{figure}

Figures \ref{fig:g21} and \ref{fig:g32} show, in the $L_\textup{X}-\Gamma_{0.2-1~{\rm keV}}$ and $L_\textup{X}-\Gamma_{0.5-2~{\rm keV}}$ plane respectively, the 60 selected sources. Figure \ref{fig:g21} shows the 31 sources selected in the ultrasoft band while Figure \ref{fig:g32} the 29 retrieved from the soft band selection.\\

   
Fifteen objects, out of the remaining 60, are well-studied sources with abundant literature. They are characterized by extreme X-ray properties, both in terms of (soft) spectral shape and variability on long and short terms: 
   \begin{itemize}
       \item eight TDEs or TDE candidates: 2MASX J02491731-0412521 \citep{esq07}, ASASSN-15oi \citep{hol16}, ASASSN-14li \citep{mak14}, 2XMM J123103.2+110648 \citep{lin17b}, 2MASS J12013602+3003052 \citep{sax12}, SDSS J150052.07+015453.8 \citep{lin17a}, GSN 069 \citep{shu18,min19,she21}, which is a convincing TDE candidate but also show huge X-ray bursts along with QPEs and slow decay in luminosity with possible rebrightening, 2MASX J19271951+6533539 (a changing look AGN with variability possibly caused by a TDE, \citealt{ric20}), 2XMM J141711.0+522541 (a likely TDE onto an IMBH appearing as a bright off-nuclear transient, \citealt{lin16});
       \item seven extreme AGN, characterized by supersoft X-ray emission coupled with outstanding short and/or long term variability: RX J1301 \citep{giu20} which is associated with QPEs, 2MASX J10343860+3938277 exhibiting a bright soft excess and quasi-periodic oscillations \citep{gie08}, the Seyfert galaxies 4U 0708-49 \citep{bol02}, 2MASX J13251937-3824524 \citep{bol97} and 2MASX J14062191+2223462 \citep{mal18} which all show highly variable narrow emission lines, and PHL 1092, a quasar with extremely variable weak lines and also often X-ray weak narrow lines \citep{min09b}.
   \end{itemize}
    The presence of these 15 sources in our sample represents a strong and reassuring check of our selection criteria and their effectiveness in retrieving supersoft sources falling into the categories described in the introduction. These 15 sources have been extensively studied and well monitored in the past in the X-ray band and therefore we will not discuss them.
   
For the 45 sources, we retrieved the EPIC data from the {\it XMM-Newton} science archive and reduced them following the standard procedure. \rev{Data from the EPIC-MOS cameras were merged together. Periods of high background were accounted for by inspecting the light curve of the entire observation and removing the interested time intervals. Source spectra were extracted from a circular region centred on the source position with $\approx15$" radius. The background spectra were extracted from \revv{source-free regions} on the same detector \revv{chip as the sources}. Due to the supersoft nature of the sources we are interested in, we consider data between 0.2 and 8 keV because most of the X-ray counts are actually concentrated in the very soft X-ray band and the spectra, in particular of the most interesting sources, are background dominated already above $\sim 5$ keV. We shall however comment on the possible impact of calibration uncertainties below 0.3 keV (see the EPIC Status of Calibration and Data Analysis\footnote{\url{https://xmmweb.esac.esa.int/docs/documents/CAL-TN-0018.pdf}} for details) on a source-by-source basis in our analysis below.} We regrouped the spectra in order to have a minimum of one count in each channel and consequently employed the C-statistic for the fitting procedure. Any further rebinning has been adopted for sole graphical purposes. The spectral analysis was performed using the XSPEC package. The standard model we adopted at first for all of our sources is  composed of an absorbed blackbody plus power-law, redshifted to the source distance ({\em zTBabs $\times$ (zbbody + zpowerlw)} in XSPEC) with an additional layer of absorption ({\em TBabs}) with column density fixed to the Galactic value.\\

 Out of these sources, we identified:
   \begin{itemize}
       \item 30 sources optically classified as broad-line AGN (BLAGN) (19 by \citealt{par18}, 4 by \citealt{ver10}, 3 by \citealt{pie16}, 2 by \citealt{rak17}, 1 by \citealt{bar17} and 1 by \citealt{esq13}). These sources are characterized by slightly steeper-than-usual X-ray spectra. Their emission is however still compatible with their optical classification and likely due to relatively strong soft-excess with black-body temperatures $kT\sim100-200$ eV with their high-energy emission taken into account by a power law. These sources, rather than being outstanding AGN, just represent the soft X-ray bright tail of the AGN population.
       \item three sources optically classified as Seyfert 2 galaxies. One of these, 2MASX J10181928+3722419 sees its emission in the soft X-ray band dominated by star-formation activity rather than accretion \citep{lam12}. NGC 6264, classified by \citealt{ver10}, hosts an H$_2$O megamaser \citep{cas13} and it is also a well-known Compton-thick AGN. The last one, 2dFGRS TGS243Z047, has been classified as a heavily obscured AGN by \citet{lac13}.
       \item two star-formation dominated sources: 2XMM J021704.5-050214 has been classified as an elusive AGN with star-forming activity \citep{men16} and 2MASX J17020882+6412210, classified as a star-forming galaxy by \citet{mic18}. The X-ray analysis of their emission confirms that their emission is in fact consistent with star-formation activity rather than nuclear activity phenomena;
       \item one source, 2dFGRS TGS431Z029, classified as a non-active galaxy \citep{lav11} given the lack of emission line in its optical spectrum, has a soft X-ray emission distinctive of an obscured, possibly Compton-thick AGN. A dedicated forthcoming publication (Sacchi et al. in preparation) will address in detail the emission by this peculiar source.
   \end{itemize}
    The spectral parameters of these 36 sources are collected in Table \ref{tab:non_int}. Figures \ref{fig:non_int1} and \ref{fig:non_int2} show the highest quality spectrum of each source. These sources will not be discussed any further.\\
   
    The remaining nine sources show X-ray emission not immediately consistent with their optical classification. To address their peculiar behaviour, these sources will be described and analyzed individually in the following Section.

\section{Peculiar Super Soft Sources}

   Here we describe in detail the sources whose X-ray properties do not resemble the typical AGN and/or that deviate from the expected behaviour based on optical classification. We modelled the X-ray emission of our sources using three different models:
   \begin{itemize}
       \item absorbed blackbody  ({\em zTBabs $\times$ zbbody} in XSPEC);
       \item absorbed powerlaw ({\em zTBabs $\times$ zpowerlw} in XSPEC);
       \item absorbed blackbody + powerlaw ({\em zTBabs $\times$ (zbbody + zpowerlw)} in XSPEC);
   \end{itemize}
   each spectral component is redshifted to the source distance, and we added an additional layer of absorption ({\em TBabs}) with column density fixed to the Galactic value for each spectral model. 
   
   When multiple observations were available for the same source, we fitted all observations jointly. We started by keeping all the fitted parameters linked together over different observations, freeing them one at a time and checking for any improvement of the statistics through an F-test with $5\%$ acceptance threshold. The resulting best-fitting models and spectral parameters are reported in Table \ref{tab:gen}. 
   
   As per X-ray light curves, we found no clear short-term variability in our sample. As a cautionary remark, it is worth pointing out that this could be partly due to the fact that the sources are generally faint, and therefore need temporal binning often larger than 1 ks in order to obtain meaningful signal-to-noise ratios in the individual bins.
   
   We divided these nine sources into three groups based on their spectral properties, as detailed below.
   
\begin{table*}
\caption{\label{tab:gen} \rev{Constraints on the spectral fits to the nine supersoft sources.}}
\centering
\begin{tabular}{|c|c||cc||ccc||ccc||c|}
\hline
\multirow{2}{*}{lab.}&\multirow{2}{*}{date}& \multicolumn{2}{c||}{blackbody} & \multicolumn{3}{c||}{powerlaw} &\multicolumn{3}{c||}{blackbody+powerlaw}&\multirow{2}{*}{$\log_{10}L$}\\
& &$C/\nu$&$kT$&$C/\nu$&$N_\textup{H}$&$\Gamma$&$C/\nu$&$kT$&$\Gamma$&\\
\hline
\multirow{3}{*}{D1}&2010-8-20&\multirow{3}{*}{\bf 155/162}&\multirow{3}{*}{$0.15_{-0.01}^{+0.01}$}&\multirow{3}{*}{179/162}&\multirow{3}{*}{\it 0.0}&\multirow{3}{*}{$3.8_{-0.2}^{+0.2}$}&\multirow{3}{*}{159/160}&\multirow{3}{*}{$0.14_{-0.02}^{+0.02}$}&\multirow{3}{*}{$3.6_{-0.7}^{+0.6}$}&$41.83_{-0.07}^{+0.07}$\\
&2013-8-17& & & & & & & & &\multirow{2}{*}{$41.43_{-0.04}^{+0.04}$}\\
&2015-5-30& & & & & & & & &\\
\hline
\multirow{3}{*}{D2}&2007-1-8&\multirow{3}{*}{281/267}&\multirow{3}{*}{$0.14_{-0.01}^{+0.01}$}&\multirow{3}{*}{\bf 279/267}&\multirow{3}{*}{\it 0.0}&\multirow{3}{*}{$5.0_{-0.2}^{+0.2}$}&\multirow{3}{*}{278/265}&\multirow{3}{*}{\it 0.12}&\multirow{3}{*}{$5.0_{-0.4}^{+0.2}$}&\multirow{2}{*}{$44.63_{-0.05}^{+0.05}$}\\
&2008-7-3& & & & & & & & &\\
&2015-2-6& & & & & & & & &$44.10_{-0.05}^{+0.05}$\\
\hline
P1&2011-5-8&104/49&$0.043_{-0.004}^{+0.004}$&106/49&{\it 0.0}&$7.3_{-0.7}^{+0.8}$&\bf{87/47}&$0.024_{-0.004}^{+0.004}$&$2.7_{-0.1}^{+0.1}$&$41.77_{-0.09}^{+0.08}$\\
\hline
\multirow{4}{*}{P2}&2004-7-7&39/51&$0.44_{-0.04}^{+0.04}$&\bf{34/50}&$0.3_{-0.1}^{+0.2}$&$2.9_{-0.4}^{+0.5}$&-&-&-&$44.8_{-0.2}^{+0.2}$\\
&2010-12-8&\multirow{3}{*}{375/158}&\multirow{3}{*}{$0.36_{-0.02}^{+0.02}$}&\multirow{3}{*}{292/158}&\multirow{3}{*}{\it 0.0}&\multirow{3}{*}{$2.6_{-0.1}^{+0.1}$}&\multirow{3}{*}{\bf 182/155}&\multirow{3}{*}{$0.02_{-0.01}^{+0.01}$}&\multirow{3}{*}{$2.2_{-0.1}^{+0.1}$}&$44.18_{-0.04}^{+0.04}$\\
&2010-12-16& & & & & & & & &\multirow{2}{*}{$44.05_{-0.05}^{+0.06}$}\\
&2010-12-18& & & & & & & & &\\
\hline
\multirow{4}{*}{P3}&2001-7-4&\multirow{4}{*}{\bf 126/140}&$0.12_{-0.01}^{+0.01}$&\multirow{4}{*}{134/140}&\multirow{4}{*}{\it 0.0}&$4.8_{-0.3}^{+0.3}$&\multirow{4}{*}{164/138}&$0.11_{-0.01}^{+0.01}$&\multirow{4}{*}{$1.5_{-0.3}^{+0.3}$}&\multirow{4}{*}{$43.83_{-0.05}^{+0.05}$}\\
&2016-7-30& &$0.044_{-0.004}^{+0.010}$& & &$9.1_{-1.6}^{+1.6}$& &$0.033_{-0.04}^{+0.03}$& & \\
&2016-8-13& &$0.24_{-0.02}^{+0.03}$& & &$3.0_{-0.3}^{+0.3}$& &\multirow{2}{*}{$0.11_{-0.01}^{+0.01}$}& & \\
&2017-1-4& &{\it 0.12}& & &{\it 3.0}& & & &\\
\hline
T1&2006-11-29&\bf{104/92}&$0.11_{-0.01}^{+0.01}$&110/91&$0.14_{-0.06}^{+0.1}$&$5.4_{-0.7}^{+1.0}$&101/89&$0.09_{-0.02}^{+0.02}$&$1.9_{-0.9}^{+1.1}$&$41.33_{-0.04}^{+0.12}$\\
\hline
T2&2018-1-6&\bf{97/95}&$0.057_{-0.005}^{+0.004}$&112/95&{\it 0.0}&$6.3_{-0.3}^{+0.3}$&124/93&$0.063_{-0.005}^{+0.004}$&$-0.9_{-1.0}^{+0.7}$&$41.57_{-0.07}^{+0.07}$\\
\hline
\multirow{2}{*}{T3}&2009-10-31&-&-&-&-&-&-&-&-&{\it 42.56}\\
&2012-12-3&\bf{51/41}&$0.08_{-0.03}^{+0.04}$&70/41&{\it 0.0}&$5.2_{-0.2}^{+0.2}$&45/39&$0.08_{-0.03}^{+0.04}$&$1.0_{-0.5}^{+0.6}$&$43.16_{-0.04}^{+0.04}$\\
\hline
\multirow{2}{*}{T4}&2010-8-8&\multirow{2}{*}{\bf{224/234}}&\multirow{2}{*}{$0.062_{-0.003}^{+0.004}$}&\multirow{2}{*}{230/234}&\multirow{2}{*}{{\it 0.0}}&\multirow{2}{*}{$6.1_{-0.3}^{+0.4}$}&\multirow{2}{*}{224/233}&\multirow{2}{*}{$0.063_{-0.003}^{+0.004}$}&\multirow{2}{*}{{\it 3.0}}&$41.04_{-0.10}^{+0.09}$\\
&2022-1-17& & & & & & & & &$39.7_{-0.2}^{+0.1}$\\
\hline
\end{tabular}
\tablefoot{The quantities in italic are fixed during the fit procedure, boldface indicates the statistically favoured model. \rev{$C$ is the value of C-statistic and $\nu$ are the degrees of freedom.} The blackbody temperature is in keV, the \rev{intrinsic absorption} Hydrogen column density is in units of $10^{20}$ atoms/cm$^2$. The \rev{intrinsic absorption} columns are omitted for the blackbody and blackbody+powerlaw models as no intrinsic absorption was required in any source. The \rev{unabsorbed} luminosity, expressed in erg/s, is computed over the $0.5-2$ keV range and adopting the best fitting model. The luminosity of the first observation of the source labelled T3 is a $3\sigma$ upper limit. \rev{The uncertainties reported correspond to
a change in the fit statistics of $\Delta C=1$.}}
\end{table*}   

\subsection{Sources with a drop in flux above 2 keV}

\begin{figure}
	\includegraphics[width=\hsize]{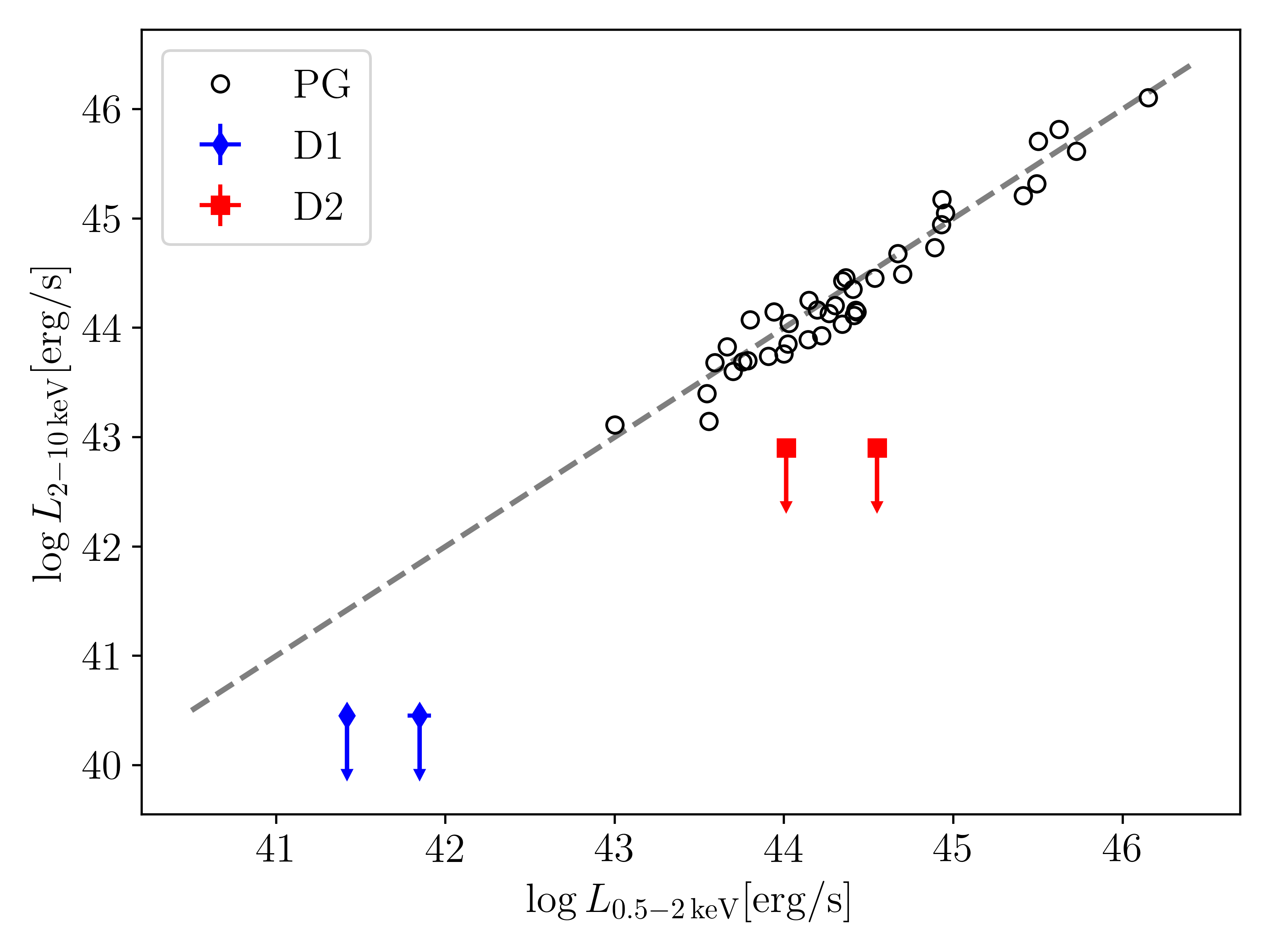}
    \caption{$L_{2-10\,\rm keV}$ vs. $L_{0.5-2\,\rm keV}$ for PG quasars (black circles) and for our sources (coloured markers). The dashed line is the bisector. For each of the two sources, the two points refer to different observations. The hard band luminosities are upper limits.}
    \label{fig:hs}
\end{figure}

   By visual inspection of our sources, we found two of them showing an extremely soft spectrum mainly because of a steep drop in their flux above $\sim2$ keV. In order to estimate the importance of the soft excess we compared the emission in the $0.5-2$ keV and $2-10$ keV bands. The emission in the hard band was calculated modelling the sole data above 2 keV with a power-law with photon index fixed to 1.9 and fixed Galactic absorption; this is done in order to avoid the fit in the hard band being dominated by the extrapolation of a power-law component that is constrained only by soft data. Note that this differs from the best-fitting parameters reported in Table \ref{tab:gen}, but is a better choice in order to isolate the contribution of the two spectral components. In both cases, the emission in the hard band is so faint that only upper limits were in fact retrieved. Figure \ref{fig:hs} shows the comparison between these supersoft sources and standard (in terms of X-ray properties) PG quasars \citep{pic05,min09a}. It is clear that the sources from our sample, have much larger soft excess than typical broad line quasars or, in other words, that they are extremely hard X-ray-weak sources, completely dominated by a thermal-like component.
   
\begin{figure}
	\includegraphics[width=\hsize]{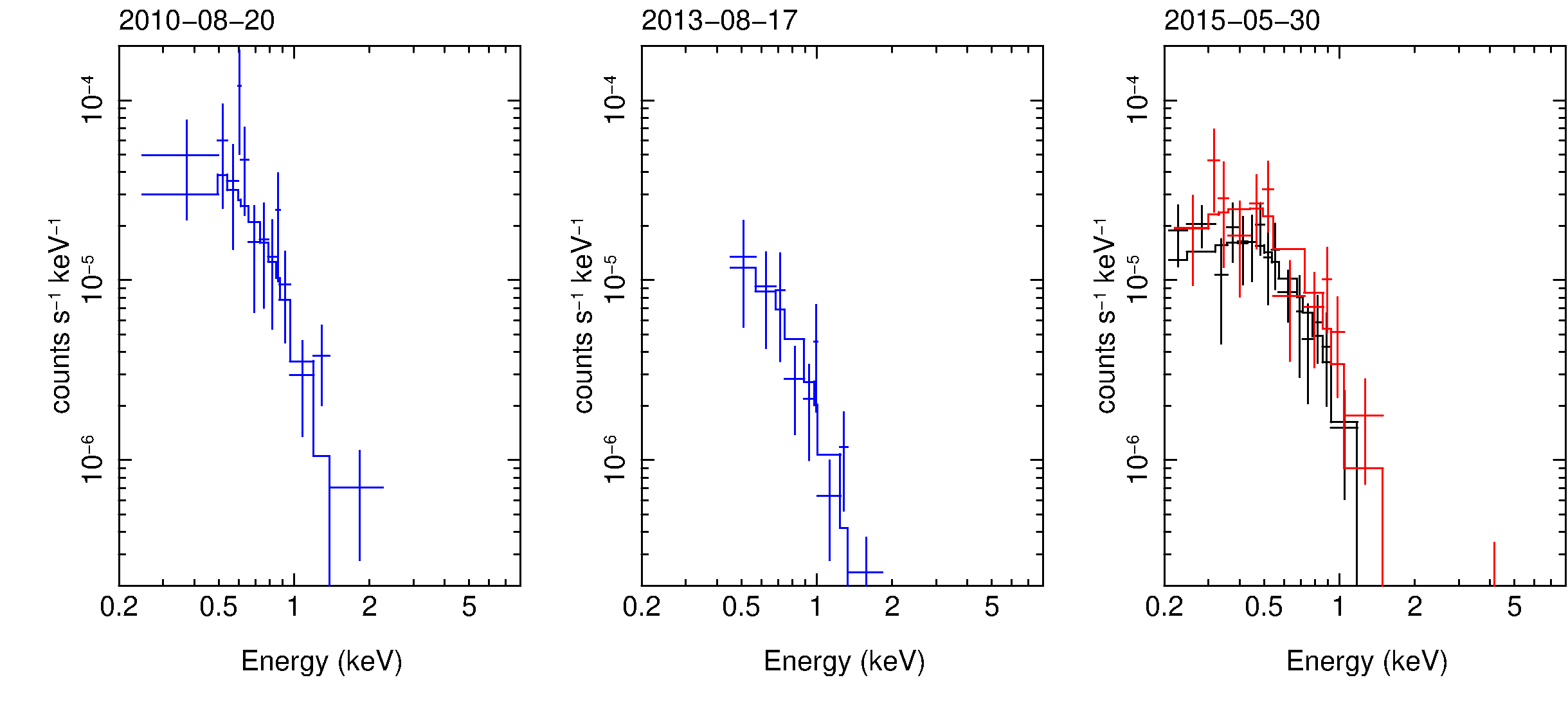}
    \caption{X-ray spectra of the source labelled D1. Blue  data are from Chandra, black from the {\it XMM-Newton} EPIC-pn camera, and red from the merged EPIC-MOS one. Solid lines show the best-fitting model.}
    \label{fig:d1}
   \end{figure}

\paragraph{[CRS2013] 4,} labelled as D1, has a photometric redshift of $z=0.115$ from the SDSS DR8 \citep{aih11} and no optical spectral information is available. Although this source is probably associated with the outskirts of the ACO 2443 cluster \citep{cla13}, its X-ray spectrum strongly deviates from thermal plasma emission (e.g. the {\em apec} model in XSPEC). This source has been observed on three occasions: twice by {\it Chandra}, in 2010 and 2013 and once by {\it XMM-Newton}, in 2015. \rev{{\it Chandra} data were extracted following the standard procedure, extraction region for source and background were taken as a circular region of $\approx2$" around the source position and as a disjointed annulus surrounding it respectively. Data were regrouped to have at least one count in each channel. Spectral fitting was performed on data between 0.3 and 8 keV.} The spectrum of the source is best fitted by a blackbody with a temperature of about 150 eV and no intrinsic absorption nor power-law component in the hard X-ray band. The source luminosity in the $0.5-2$ keV range shows significant variability on years timescales, going from $6.5$ to $2.5\times10^{42}$ erg/s from the first to the second observation, while it remains constant from the second to the third.  The soft X-ray emission of D1 is at least one order of magnitude larger than in PG quasars with respect to that expected from the hard X-ray luminosity upper limit of $3\times10^{40}$ erg/s. Its X-ray spectrum is shown in Figure \ref{fig:d1}.
   
\begin{figure}
	\includegraphics[width=\hsize]{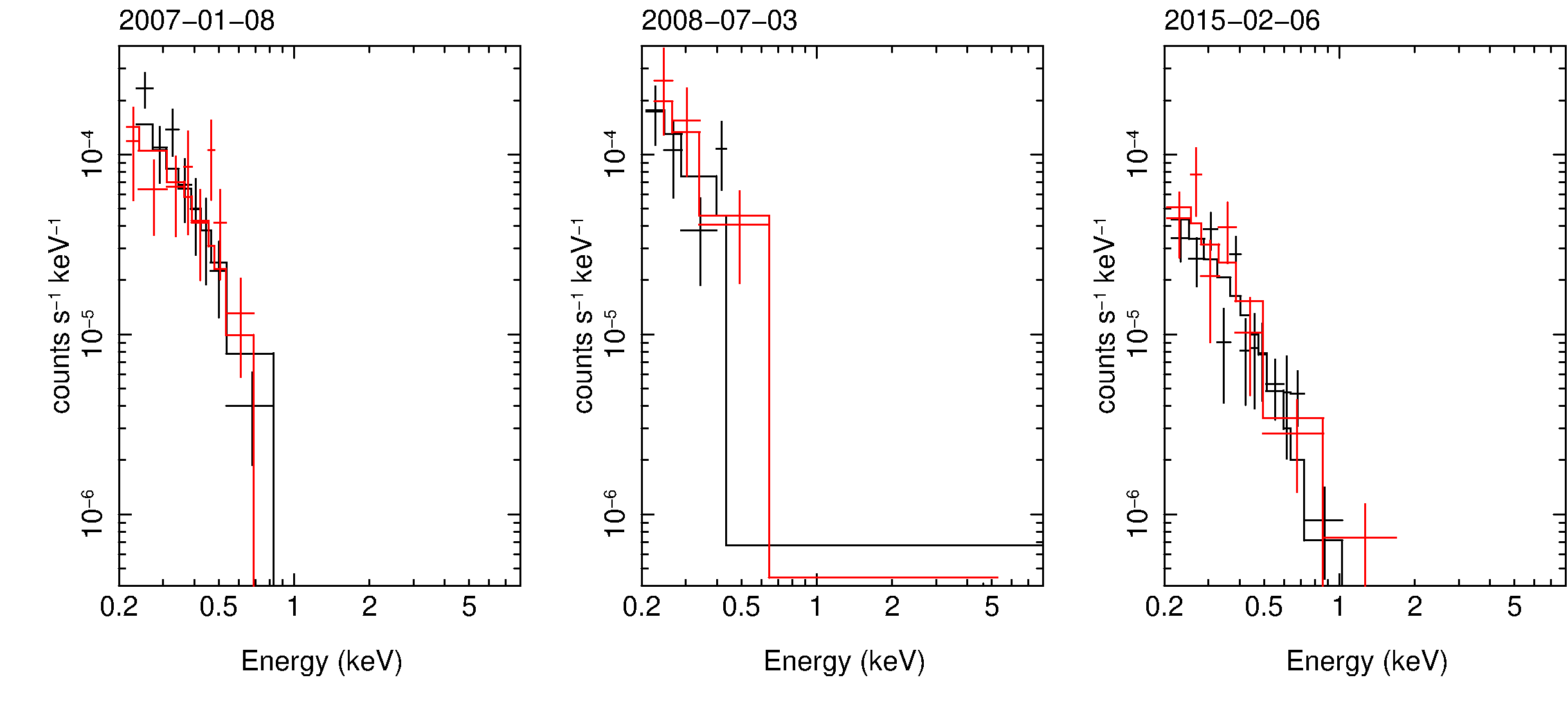}
    \caption{X-ray spectra of the source labelled D2. The colour code is the same as in the previous plots. We only show the highest signal-to-noise observation of the three performed in July 2008.}
    \label{fig:d2}
   \end{figure}

\paragraph{2XLSSd J021728.4-041346,} labelled as D2, is located at $z=1.173$ and catalogued as an AGN by spectral energy distribution (SED) fitting \citep{mel13,liu16}, as only photometric information is available. The source has five {\it XMM-Newton} observations, the first three were taken between  the 8th and the 9th of January 2007 and are here merged together. The fourth observation was taken in July 2008, and the last one has been taken in February 2015. The data of all observations were fitted jointly and the best fitting model is an extremely steep power law with photon index $\Gamma=5$ (or, almost equivalently, a blackbody with a temperature of about 140 eV). The X-ray luminosity on the 0.5-2 keV range of the source suffers a drop of almost a factor 3.5 from $4\times10^{44}$ to $1.2\times10^{44}$ erg/s between the first and last observation. In each observation, the soft X-ray emission is more than one order of magnitude larger in comparison to the standard PG quasars given that its $2-10$ keV emission sits below $8\times10^{42}$ erg/s, as shown in Figure \ref{fig:hs}. The X-ray spectrum of the source is shown in Figure \ref{fig:d2}.  
   
  \subsection{Peculiar AGN}

  Three sources of our sample are optically classified as AGN and are selected here because they exhibit a peculiar supersoft spectral component at least during one observation. As for the previously discussed sources, no clear short-term variability has been found in these sources.
  
\begin{figure}
        \centering
	\includegraphics[width=0.5\hsize]{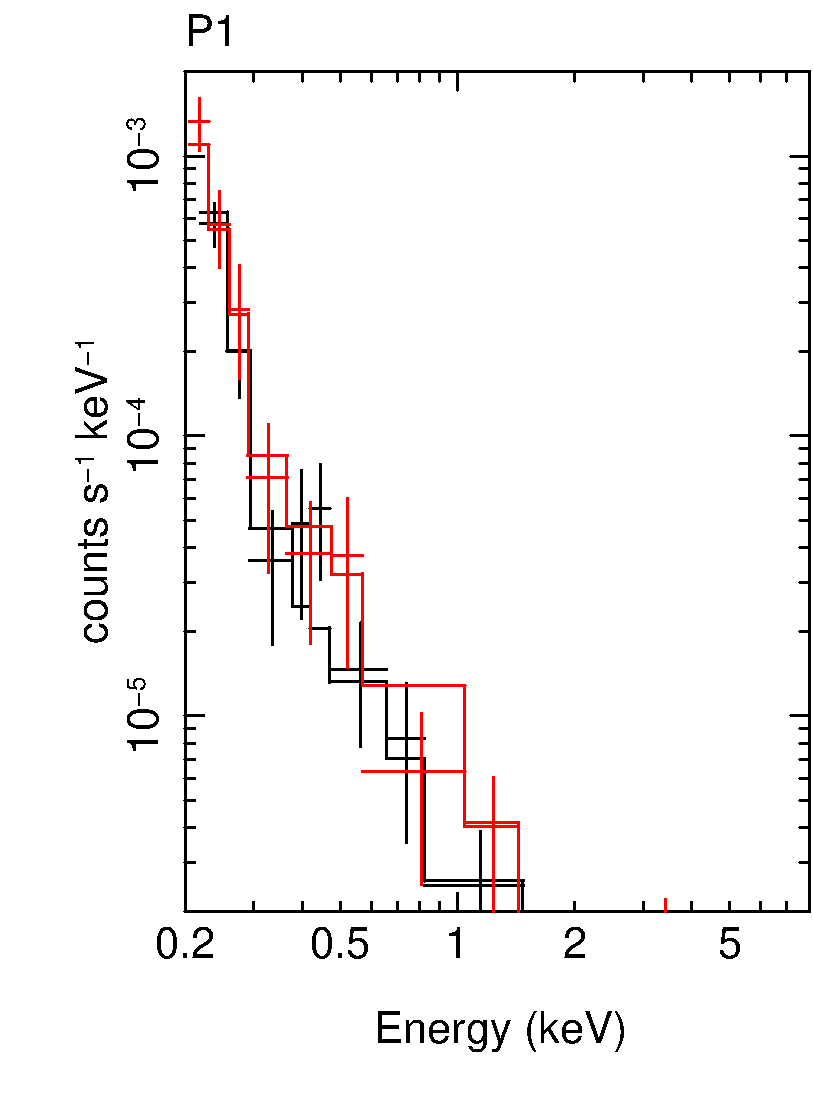}
    \caption{X-ray spectrum of source labelled P1. The colour code is the same as in the previous plot.}
    \label{fig:p1}
   \end{figure}

\paragraph{[VV2006] J231419.8-525901,} labelled as P1, is located at a spectroscopic redshift $z=0.15575$ \citep{jon09} and classified as a Seyfert 1 galaxy \citep{ver06}. It has been observed once by {\it XMM-Newton} and its spectrum is best fitted by a blackbody plus power-law model with no intrinsic absorption. The power law is steep and mostly constrained by soft X-ray data below 2 keV, with a photon index $\Gamma>2.6$. What makes this source really peculiar is the extremely cold blackbody of just $kT\sim20$ eV dominating the ultrasoft X-ray band. Such a cold thermal-like component, detected in both the EPIC-pn and MOS cameras, is unprecedented in the X-ray spectra of type 1 AGN. The X-ray luminosity of P1 is about $6\times10^{41}$ erg/s. The X-ray spectrum of the source is shown in Figure \ref{fig:p1}.
   
\begin{figure}
	\includegraphics[width=\hsize]{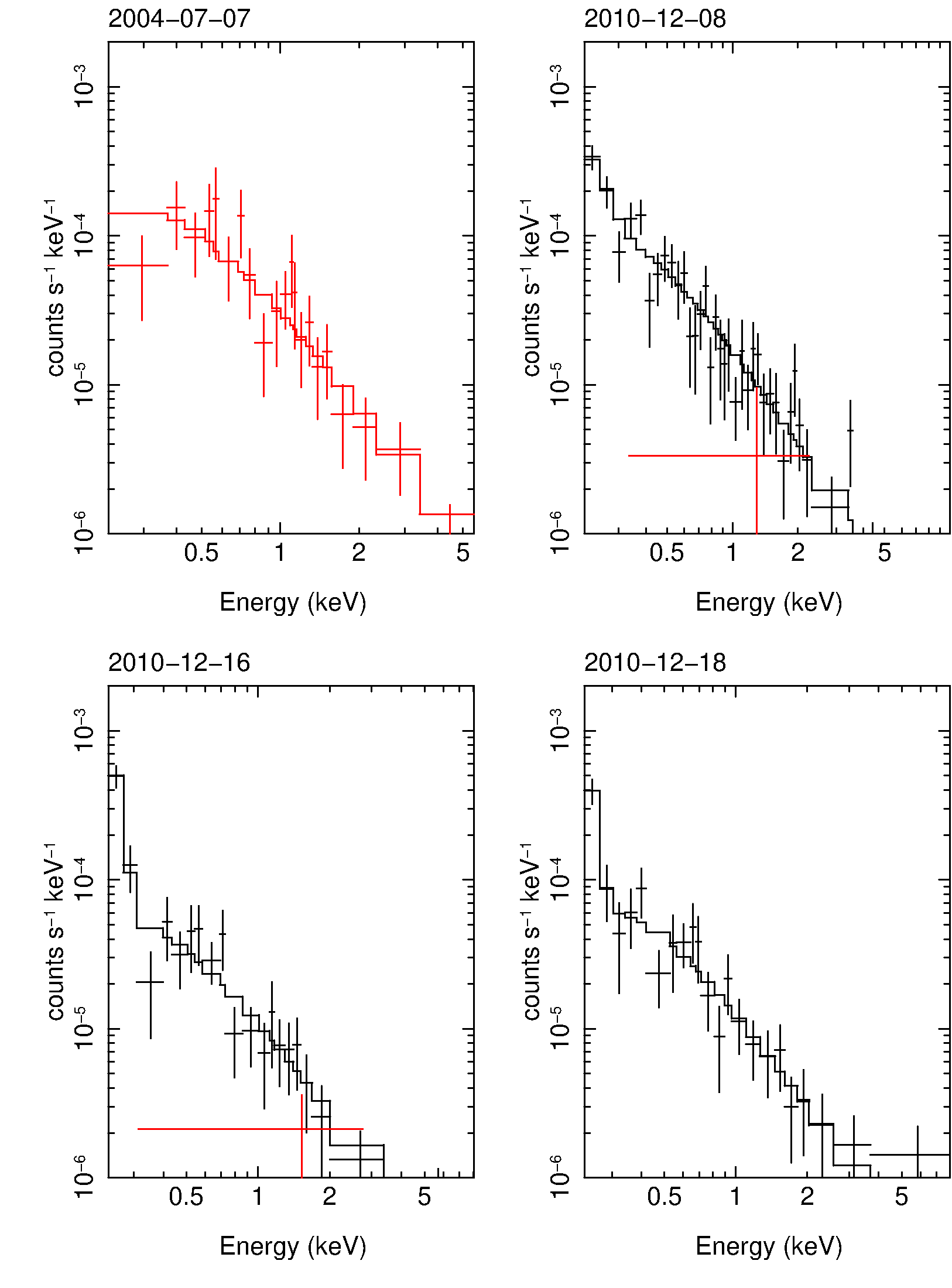}
    \caption{X-ray spectra of the source labelled P2. The colour code is the same as in the previous plots.}
    \label{fig:p2}
   \end{figure}

\paragraph{2XMM J122517.6+175726,} labelled as P2, has a spectroscopic redshift $z=0.8053$ and it is catalogued as a QSO \citep{par14}. It has been observed once in July 2004 and then three times on December 2010. For the first observation, we have data only from the EPIC-MOS camera, while for the last three observations we have only EPIC-pn data. We analyzed the first observation separately while the last three were treated together as they were taken only a few days apart from one another. In 2004, given the poor quality of the data, the source spectra are best fitted by a simple power-law model ($\Gamma\sim3$) with a drop in flux below $\sim0.4$ keV. We modelled this drop with extra-absorption at the source redshift which results in a column density of $2.7^{+1.8}_{-1.5}\times10^{21}$ atoms/cm$^2$. The unabsorbed luminosity of the source in the first observation is about $6\times10^{44}$ erg/s, consistent with the QSO classification. Remarkably, no extra-absorption is present during the 2010 observations, and the spectrum of the source is best fitted by a blackbody plus power-law model, with a very low blackbody  temperature of $kT\sim20$ eV, and the photon index of the power-law $\Gamma\sim2$. The appearance of such a cold thermal-like component makes P2 a similar source to P1 in terms of spectral shape. The luminosity of the source in 2010 is about $1.5\times10^{44}$ erg/s and shows a small but statistically significant decrease between the first and last two observations. The X-ray spectrum of the four observations is shown in Figure \ref{fig:p2}. We must point out that the supersoft component is only detected in observation where only EPIC-pn data are available. We cannot rule out that it is spurious and due to calibration uncertainties. This caveat does not fully apply to source P1 as the supersoft component is consistently detected by both EPIC-pn and EPIC-MOS cameras.
   
\begin{figure}
	\includegraphics[width=\hsize]{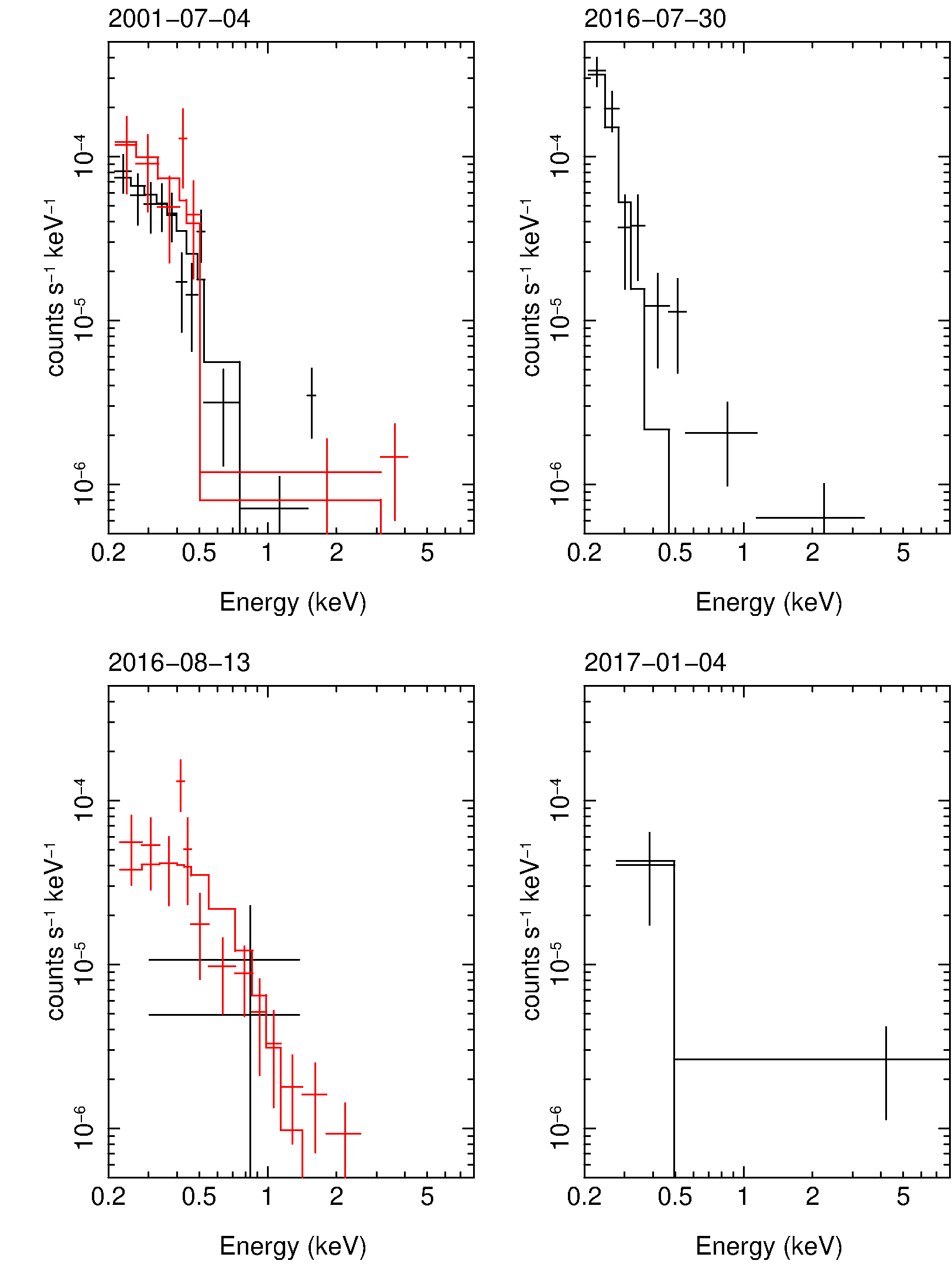}
    \caption{X-ray spectra of the source labelled P3. The colour code is the same as in the previous plots.}
    \label{fig:p3}
   \end{figure}

\paragraph{2XMM J022255.9-051352,} labelled as P3, is catalogued as a Seyfert 1 galaxy \citep{ver10} and has a spectroscopic redshift $z=0.846478$. The source was observed on four different occasions (once in 2001, two times in 2016 and one final time in 2017, although in this last observation the quality of the data is extremely poor) and its X-ray spectrum shows strong spectral variability associated with roughly constant luminosity, although the same caveat highlighted for source P2 is to be kept in mind. The best-fitting model for this source is a blackbody with no intrinsic absorption. During the first 2016 observation, the spectrum is dominated by a cold blackbody emission with $kT\sim40$ eV, while during the  second (2 weeks later) the temperature is a much higher $kT\sim240$ eV. In 2001 and 2017 (this last observation has a much lower signal-to-noise), the blackbody temperature is instead $kT\sim120$ eV. Unfortunately, as the supersoft component is only detected in the EPIC-pn data, we can not rule out the possibility of said component being related to time-dependent calibration uncertainties. The X-ray spectrum of the source is shown in Figure \ref{fig:p3}.
   
\subsection{Candidate TDEs}

   Here we present a detailed analysis of four sources that we consider strong TDE candidates although other scenarios cannot be completely ruled out. Two of these sources (T1 and T2), classified as non-active, star-forming galaxies owing to their optical properties, show soft X-ray emission, characterized by high luminosity and a steep spectral shape, which strongly favours a TDE interpretation. Unfortunately, these two sources have only one observation, preventing us to study their long-term evolution. The other two sources (T3 and T4), instead, have been observed more than once and we can therefore provide long-term information, coupled with X-ray spectroscopy and optical classification. Below we show the properties of each of these TDE candidates.

\begin{figure}
	\includegraphics[width=\hsize]{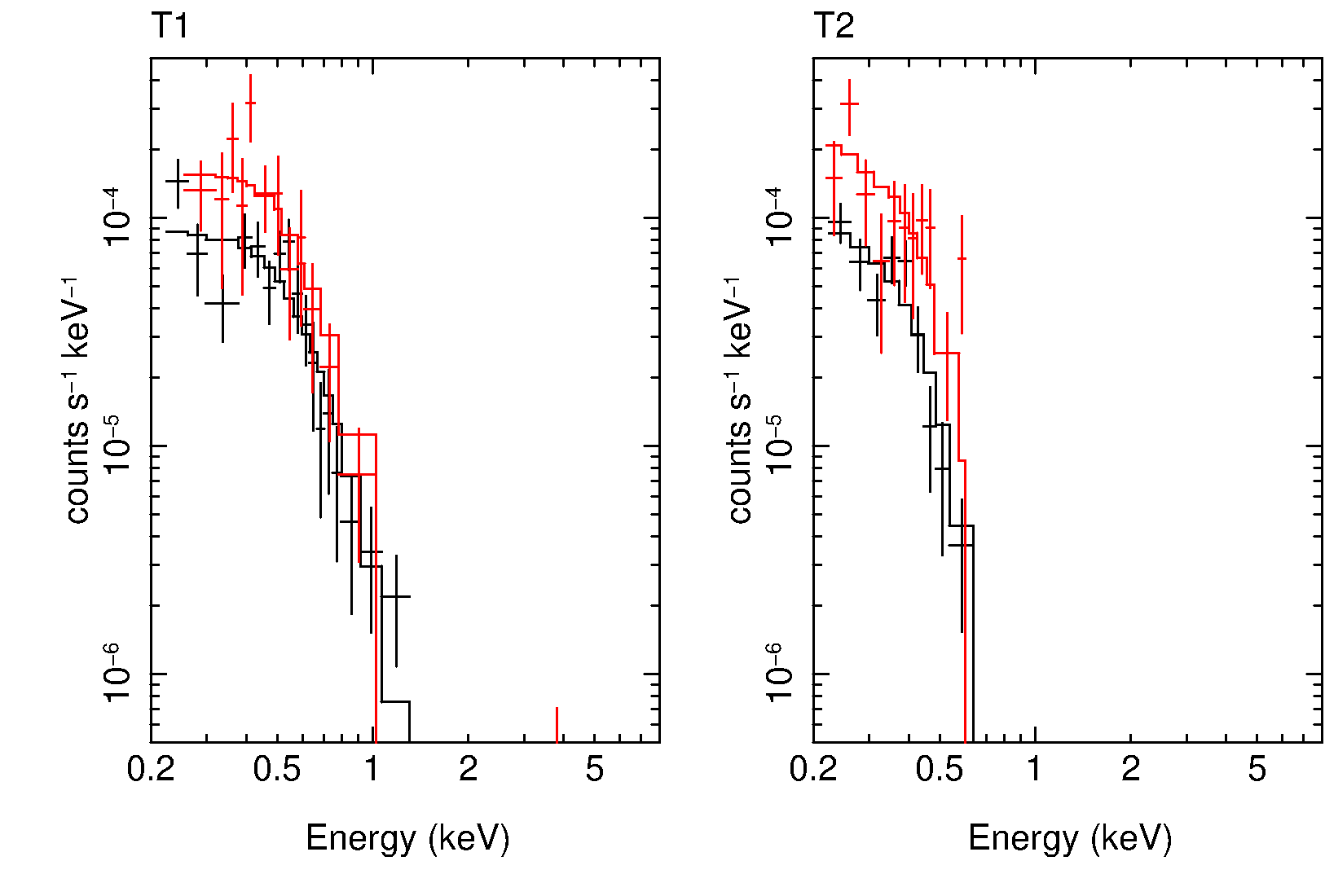}
    \caption{X-ray spectra of the candidate TDEs with only one observation, T1 (left panel) and T2 (right panel). The colour code is the same as in the previous plots.}
    \label{fig:t1t2}
   \end{figure}

   \paragraph{2MASX J00414632-2827423,} labelled as T1, is located at a spectroscopic redshift $z=0.0744$ \citep{col01}. It is classified as a non-active galaxy \citep{pat03}. In its sole {\it XMM-Newton} observation the spectrum of this source can be well reproduced by a cold blackbody of about 100 eV temperature ($kT=0.11\pm0.01$ keV). The X-ray luminosity of this source is about $2\times10^{41}$ erg/s. Although a TDE scenario can well explain the properties of this source, the source is in the far outskirts of the Sculptor cluster and hence its X-ray emission could be in principle associated with the diffuse cluster emission. \rev{In order to check this hypothesis, we extracted background spectra from different source-free regions (always on the same detector's chip), but no differences in the spectral fitting results were found. Moreover, the statistical quality of fits made using a thermal plasma emission model (as expected from the cluster's hot gas) is significantly worse than that obtained with the blackbody model ($\Delta C=21.02$ with the same degrees of freedom), so that the source appears likely accretion-powered.} The X-ray spectrum of the source is shown in Figure \ref{fig:t1t2}.

\paragraph{GAMA 91637,} labelled as T2, has a spectroscopic redshift $z=0.1775$ from the GAMA survey \citep{dri11}. This source is optically classified as a non-active galaxy \citep{lis15}. It has been observed only once with {\it XMM-Newton} and its X-ray spectrum is best fitted by a cold blackbody ($kT\sim60$ eV) with no intrinsic absorption. We tested once again the blackbody model versus the thermal plasma model. Again we found that a blackbody can describe data significantly better than the rival thermal plasma model ($\Delta C=5.88$ for the same degrees of freedom). The X-ray luminosity for this source is $\sim4\times10^{41}$ erg/s, and its X-ray spectrum is shown in Figure \ref{fig:t1t2}.

\begin{figure}
	\includegraphics[width=\hsize]{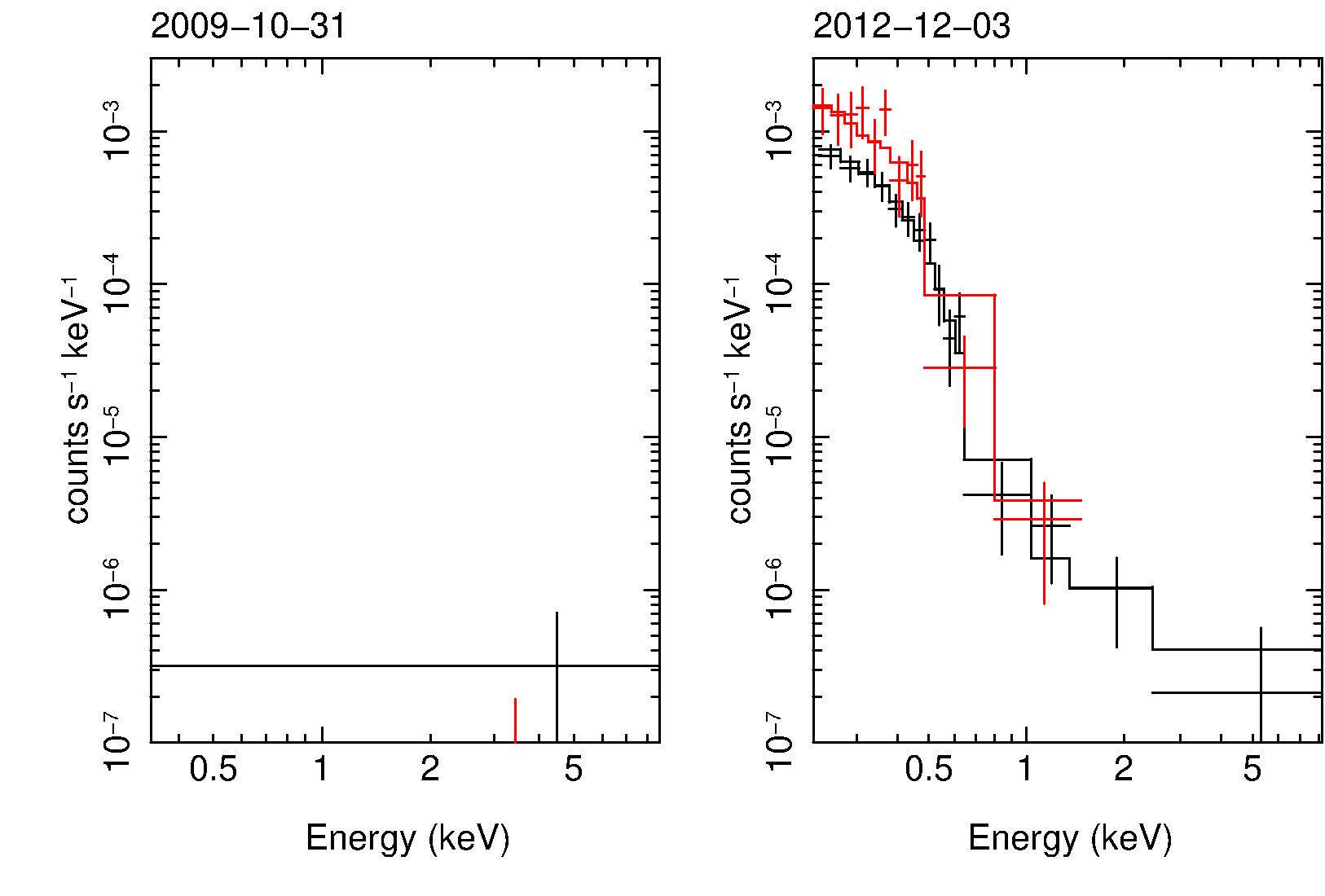}
    \caption{X-ray spectra of the source labelled T3. The colour code is the same as in the previous plots.}
    \label{fig:t3}
   \end{figure}

\paragraph{XXL-AAOmega J234255.95-543001.8,} labelled as T3, has a spectroscopic redshift $z=0.28633$ and, due to the presence of broad emission lines in its optical spectrum, it is catalogued as AGN \citep{pie16,lid16}. It has been observed with {\it XMM-Newton} once in October 2009 and a second time on December 2012. During the first observation, the source was not detected and the $3\sigma$ upper-limit to its luminosity is $\sim 4\times10^{42}$ erg/s. \rev{This upper limit has been computed using the upper limit server\footnote{\url{http://xmmuls.esac.esa.int/upperlimitserver/}} assuming a $kT=100$ eV blackbody spectrum.} During the second observation, the luminosity of the source was $\sim1.5\times10^{43}$ erg/s, almost an order of magnitude higher than the upper limit of three years earlier. A blackbody with a temperature $kT\sim 80$ eV and no intrinsic absorption can very well reproduce the spectrum during this second observation. These X-ray properties appear to be consistent with a TDE detected near its peak, possibly occurring in a pre-existing faint AGN. An alternative scenario involves an extreme variability event from an AGN. The X-ray spectrum of the source is shown in Figure \ref{fig:t3}.
   
\paragraph{SDSS J152717.95+164503.2,} labelled as T4, has a spectroscopic redshift $z=0.0606$ from the SDSS DR9 \citep{ahn12}. This source, as T1 and T2, is catalogued as a non-active, star-forming galaxy with a star-formation rate of $\log {\rm SFR}=-0.1192\,M_\odot/$year \citep{tob14,dua17}. It had only one archival observation with {\it XMM-Newton}, in August 2010. At that epoch, the X-ray spectrum of the source can be well modelled by a cold blackbody with $kT\sim60$ eV and no intrinsic absorption. Given the optical classification of star-forming galaxies, we challenged the blackbody model with a thermal plasma model ({\em apec} model in XSPEC), in order to rule out the possibility that the soft X-ray emission is due to star formation rather than accretion. We found that the blackbody model reproduces the data significantly better ($\Delta C=5.36$ with the same degrees of freedom). The X-ray luminosity between 0.5 and 2 keV of the source was $10^{41}$ erg/s. 
   
\subsection{Follow-up observation of T4}

\begin{figure}
	\includegraphics[width=\hsize]{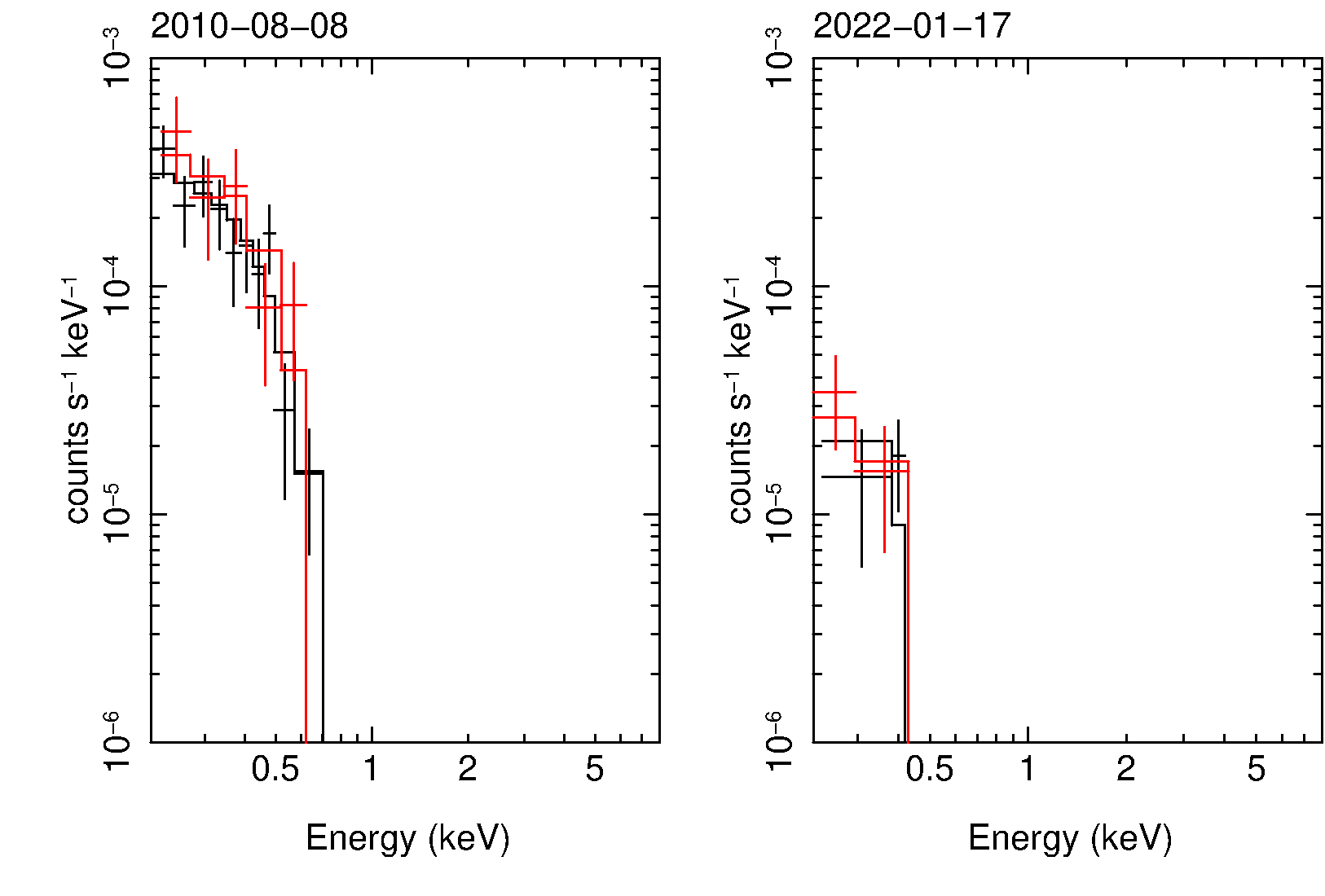}
    \caption{Left panel: archival X-ray spectrum of the source labelled T4 from an observation in 2010. Right panel: the new X-ray spectrum obtained with the new pointed observation in 2022. The soft flux dropped by a factor of $\sim20$. The colour code is the same as in the previous plots.}
    \label{fig:t4}
   \end{figure}

One obvious way to confirm the TDE nature of the four candidates is through a new X-ray observation: the long-term decay of supersoft sources has been used for decades to assess its TDE nature. We asked for new observations with {\it XMM-Newton} and the source labelled as T4 was re-observed in January 2022. After 12 years the spectral shape of the source is unchanged, strongly suggesting that we are still observing the source rather than the host galaxy emission, and its X-ray luminosity dropped by a factor $\approx20$. Its spectrum is shown in Figure \ref{fig:t4}.

\section{Discussion}
   
   We presented a sample of supersoft, luminous X-ray sources that are associated with the central regions of galaxies with the main goal of identifying a series of potentially interesting objects ranging from overlooked X-ray TDEs to supersoft AGN.  Starting from the more than $8\times10^5$ detections of the latest {\it XMM-Newton} catalogue, we obtained a final sample of 60 sources sharing similar spectral features and X-ray luminosities well above $10^{41}$ erg/s. Out of these 60 sources, 15 turned out to be well-known TDEs, TDE candidates, or AGN with extreme spectral/variability properties, confirming that our selection criteria indeed can single out the correct population of X-ray sources. By performing a complete X-ray spectral analysis of the remaining 45 sources, we identified 36 objects that are either standard type 1 AGN (although somewhat softer-than-usual in the X-ray) or sources such as obscured AGN and/or star-forming galaxies where the soft X-ray emission is dominated by extended emission. One source, 2dFGRS TGS431Z029 is somewhat peculiar as it exhibits an X-ray spectrum consistent with a heavily obscured AGN despite being classified as a non-active galaxy due to the lack of optical emission lines (Sacchi et al. in preparation). The remaining nine sources have spectral properties that most significantly deviate from the expected behaviour based on optical classification and/or do not resemble those of standard AGN. We focus our discussion on these nine sources.
   
   The X-ray spectrum of two sources (D1 and D2) is best described by a blackbody model with a temperature of 140-150 eV (or a steep power-law with $\Gamma\sim 5$), rather typical of AGN soft excesses. However, their hard X-ray emission above 2 keV is much fainter than in standard AGN as only upper limits could be derived on the 2-10 keV luminosity, making D1 and D2 two examples of hard X-ray weak sources (or extreme soft excess ones). Long-term variability suggests that their soft X-ray spectrum is not dominated by extended emission nor star formation and that the lack of hard X-ray emission is unlikely to be the signature of Compton-thick AGN. On the other hand, no short-term variability was found in the X-ray during any of the observations. 
   
   Although only photometric information is available for both D1 and D2, the latter is classified as AGN by SED-fitting, and its X-ray luminosity of a few times $10^{44}$ erg/s (about two orders of magnitude higher than in D1) is consistent with that classification, although such X-ray luminosity is also consistent with the early stages of TDEs evolution. As AGN, the X-ray spectrum of D2 is reminiscent of Narrow Line Seyfert 1 galaxies, often characterized by a prominent soft X-ray excess, and could then represent a particularly extreme version of standard BH accretion, most likely associated with a high accretion rate. 
   
   The same interpretation applies to D1 as well. The possibility that D1 (and/or D2) is a TDE observed by chance relatively close to its X-ray peak appears unlikely, because of the persistent X-ray emission with only limited dimming over the course of several years (about five and eight years respectively), as well as because the inferred temperature of the best-fitting blackbody models is significantly higher than in most other X-ray TDEs. However, some cases of extremely long-lived TDE candidates characterized by higher-than-standard blackbody temperature exist (e.g. \citealt{lin17b}), so that a TDE origin for D1 (and/or D2) cannot be firmly ruled out. On the other hand, the X-ray spectrum of D1 is very similar to that of ULSs where the dominant soft X-ray emission can be understood in terms of optically thick outflows blocking the hard X-ray and reprocessing them into a soft thermal component. The X-ray luminosity in excess of $10^{42}$ erg/s is much higher than that of typical ULSs and could suggest accretion onto an IMBH. Interestingly, the most convincing IMBH candidate  to date, ESO 243-49 HLX-1 \citep{far09}, is both soft and bright enough to satisfy our search criteria, but was not selected here because of its off-centre location.  
   
   Three further sources are classified as peculiar AGN (P1, P2, and P3). They are all spectroscopically classified as AGN (two Seyfert 1 galaxies, and one QSO), and what makes them special is the presence of at least one epoch in which an extremely soft X-ray component is observed in the spectrum. When modelled as a simple blackbody, we derive temperatures as low as 20 eV in P1 and P2, and 40 eV in P3 for this cold component. It should be pointed out that the cold component is only detected by the EPIC-pn camera in P2 and P3 as the MOS data were not available at that particular epoch. This could in principle signal that the very soft emission is in fact associated with calibration uncertainties of the EPIC-pn \rev{(see the EPIC Status of Calibration and Data Analysis\footnote{\url{https://xmmweb.esac.esa.int/docs/documents/CAL-TN-0018.pdf}} for details)}. However, both EPIC cameras are consistent with each other in the case of the only observation of P1 where a rather clear $\sim 20$ eV thermal component is present together with a steep power law. 
   
   P2 is the only source in our sample where cold X-ray absorption is detected (in excess of the Galactic one). This is, however, limited to one observation only. The cold component appears six years later and is associated with unabsorbed X-ray spectra, suggesting the possibility that it may have been present always, but masked absorption during the first observation. On the other hand, the cold component in P3 is transiently seen in one observation of a source that appears to vary wildly spectrally as the best-fitting temperature of simple blackbody models is seen to vary by more than a factor of 2 in about 15 years. 
  
   Summarizing, the five sources D1, D2, P1, P2 and P3 are likely extreme examples of supersoft state in AGN, even if in some cases we can not rule out a TDE or accretion on an IMBH. Considering their rarity  (our selection narrowed down a few such sources from a parent sample of hundreds of thousands of sources), it is possible that they can be explained as the extreme tail of  the standard soft-excess emission in AGN.  Alternatively, they could be associated with some unusually rare phenomenon different from standard accretion. Further observations, both at X-ray and optical wavelengths are needed to investigate better this open problem.
 
   The final four sources have properties that make them likely TDE candidates. Three of them (T1, T2, and T4) are spectroscopically classified as non-active galaxies and, in the available {\it XMM-Newton} observations, they exhibit a soft, thermal-like X-ray spectrum that is typical of X-ray TDEs observed close to their peak. When modelled as a blackbody, we derive temperatures of 60 to 110 eV, with luminosities in the range of $10^{41}$ erg/s to a few $10^{42}$ erg/s, rather typical for TDEs. The addition of a hard power-law component did not improve the spectral description in any case. This is consistent with TDE X-ray spectral properties, especially during the early stages \citep{sax20}. As only one observation was available for T1 and T2 (we also checked for archival observations of other instruments, such as {\em Swift} and {\it ROSAT}, but only upper limits above some $10^{43}$ erg/s were available), long-term variability cannot be used to support the TDE classification, but the non-active nature of the host galaxies coupled with the relatively high X-ray luminosities and spectral shape is highly suggestive. As already discussed for D1, these three sources may be instead associated with the very high luminosity end of the ULS population. If so, they would be potential IMBH candidates that escaped detection so far because searches for ULXs have always concentrated on off-centre sources to avoid confusion with AGN.
   
   For what concern T4, its follow-up observation with {\it XMM-Newton} revealed a flux drop of a factor $\sim20$ in the soft X-ray, strongly supporting the TDE interpretation, \rev{although we observe no hints of the spectral hardening which is often observed in the later phases of TDEs}. All in all though, it provides a robust validation of our selection method, indirectly supporting the TDE classification of T1, T2 and T3 as well.

   One of the sources that we suggest as a possible TDE candidate (T3) is classified as AGN, due to the presence of broad emission lines in the optical spectrum. T3 has been observed twice by XMM-Newton.  It was not detected during the first pointing while it exhibits a rather typical TDE X-ray spectrum in the second one performed about three years later. When detected, the X-ray spectrum is best modelled by a blackbody with a temperature of 80 eV and a rather high luminosity exceeding $10^{43}$ erg/s in the soft 0.5-2 keV band. Based on the optical classification and X-ray properties, the most appealing interpretation for the behaviour of T3 is that of a TDE occurring in a pre-existing AGN. On the other hand, as the upper limit during the first observation is not tight, we cannot firmly exclude that T3 is an AGN with very weak 2-10 keV emission (if any) and highly variable, particularly cold soft X-ray excess.
   
   The presence in our initial sample, comprising 60 sources, of nine well-known TDEs or strong TDE candidates supports our ability to detect them quite efficiently. In fact, this is the lower limit of our efficiency: our algorithm selected three more TDEs, OGLE16aaa \citep{zha16}, 3XMM J152130.7+074916 \citep{lin15} and 2XMMi J1847 \citep{lin11}, which satisfy our criterion but unfortunately are not associated with their host galaxy redshift in optical catalogues; other two TDEs instead, [GHC2009] D23H-1 \citep{gez09} and PS1-13jw \citep{kan17}, were not selected because their X-ray signal-to-noise ratio is below the chosen threshold.
   
   Assuming that the four sources discussed above (T1-T4) are indeed associated with TDEs, we can attempt a rough estimate of the TDE rate and check its consistency with other results in the literature obtained with different methods. From our data, we retrieve (see appendix for details) a rate of $(2.4\pm0.2)\times10^{-5}$ gal$^{-1}$ yr$^{-1}$ which falls roughly in the middle of the range of TDE rates deduced from X-ray data  \citep{don02,esq08,kha14}.

\section{Conclusions}

  In this paper we described the selection and properties of a sample of supersoft ($\Gamma>3$), luminous ($L>10^{41}$ erg/s) X-ray sources in the nuclear region of external galaxies by cross-matching the latest releases of {\it XMM-Newton} and optical catalogues of galaxies. 
  
  Our work resulted in the selection of 60 sources, 15 of which turned out to be known examples of the type of sources we are searching for in our study: TDEs, TDE candidates or supersoft AGN with often extreme X-ray properties. Of the remaining 45 sources, 36 are likely standard AGN and/or star-forming regions based on their X-ray spectral properties, while nine have been selected as further examples of potentially interesting supersoft and X-ray luminous nuclear sources. 
  
  Out of these nine sources, five are extremely soft X-ray-dominated AGN, while the remaining four are most likely newly discovered TDEs. These results confirm that a soft X-ray spectroscopic selection is a powerful and promising way to discover new TDEs and peculiar, soft X-ray-dominated sources.

\begin{acknowledgements}
We thank Emanuele Nardini and Paolo Esposito for the priceless help, insightful discussion and precious comments.
Based on observations obtained with {\it XMM-Newton}, an ESA science mission with instruments and contributions directly funded by ESA Member States and NASA.
This publication is part of the R\&D\&I project with reference RTI2018-096686-B-C21 , funded by MCIN/ AEI/10.13039/501100011033/ FEDER Una manera de hacer Europa.
\end{acknowledgements}  

\bibliographystyle{aa} 
\bibliography{biblio} 

\appendix
\section{Calculation of TDE rate}
In order to give an estimate of the rate of TDEs based on the 4 we found with our methods, we first need to estimate in how many galaxies we would have been able to identify one if it occurred. To do so, we started by retrieving the list of {\it XMM-Newton} pointings, along with the duration of each observation. Then we cross-matched the list with the four aforementioned galaxy catalogues, paying attention not to double-count the galaxies that are included in more than one catalogue. We selected all of the galaxies within a radius of 14 arcmins from the {\it XMM-Newton} pointing, to account for the {\it XMM-Newton} field of view.
   
For each galaxy, we then computed the minimum observable flux. We did so by taking the exposure time, decreasing it by a 30\% in order to account for the background flaring, computing the correspondent $7\sigma$ sensitivity (adapting the $5\sigma$ sensitivity from \citealt{wat01}) and \rev{finally multiplying for the vignetting factor taken from Fig. 13 of the "{\it XMM-Newton} Users Handbook”, Issue 2.17, 2019 (ESA: {\it XMM-Newton} SOC).} Finally, for each galaxy we converted the minimum observable flux in a luminosity threshold ($L_\textup{th}$).
   
As in looking for supersoft hyper-luminous sources, we selected only sources with luminosity higher than a chosen threshold ($10^{41}$ erg/s). We now face a double problem: as the lightcurves of TDEs decrease with time, if the luminosity threshold of a galaxy is larger then $10^{41}$ erg/s, we would catch a TDE occurring in it only in the fraction of its lifetime its luminosity is larger then the galaxy luminosity threshold; furthermore, after enough time passed we would not be able to catch it in any galaxies at all. Each galaxy was then counted, weighted with weight $w$ computed as follows:
\begin{equation}
\begin{cases}
    w = 1\hfill {\rm if}\,L_\textup{th}<10^{41}\,{\rm erg/s}\\
    w = \left(\frac{L_\textup{th}}{10^{41}\,{\rm erg/s}}\right)^{1/k}\hfill {\rm if}\,L_\textup{th}>10^{41}\,{\rm erg/s}
\end{cases}.
\end{equation}
The last expression equals the fraction of time it takes a TDE with lightcurve slope $k$ to decrease from its peak luminosity to $L_\textup{th}$ with respect to decreasing to $10^{41}$ erg/s. Assuming a classical $-5/3$ value for the value of the lightcurve slope, the number of galaxies in which we would have observed a TDE is 55498. Given that we could not keep under control every source of uncertainty, we approximate this number to 55000 and assume a conservative error of 10\% on said value. 
   
Finally, we need to assume a value for the visibility window of TDEs i.e. the amount of time a TDE has luminosity larger than our search threshold of $10^{41}$ erg/s; in this case, we chose a six years period. This choice also roughly corresponds to the amount of time a TDE at $z=0.1$, with a peak-luminosity of $10^{44}$ erg/s, following a $t^{-5/3}$ light-curve with a characteristic time of one month, would remain detectable by {\it XMM-Newton} with a $7\sigma$ sensitivity. 
This gives us a rate of $(1.2\pm0.2)\times10^{-5}$ TDEs per galaxy per year. This rate refers only to TDEs presenting a supersoft X-ray emission (thermal TDEs), we can correct for this effect by dividing our estimated rate by the fraction of thermal TDEs in the {\it XMM-Newton} catalogue which amounts to roughly $1/2$ retrieving a final estimate for the TDEs rate of $(2.4\pm0.2)\times10^{-5}$ gal$^{-1}$ yr$^{-1}$.
   
\section{Spectral parameters}
\begin{figure*}
    \centering
	\includegraphics[width=0.9\hsize]{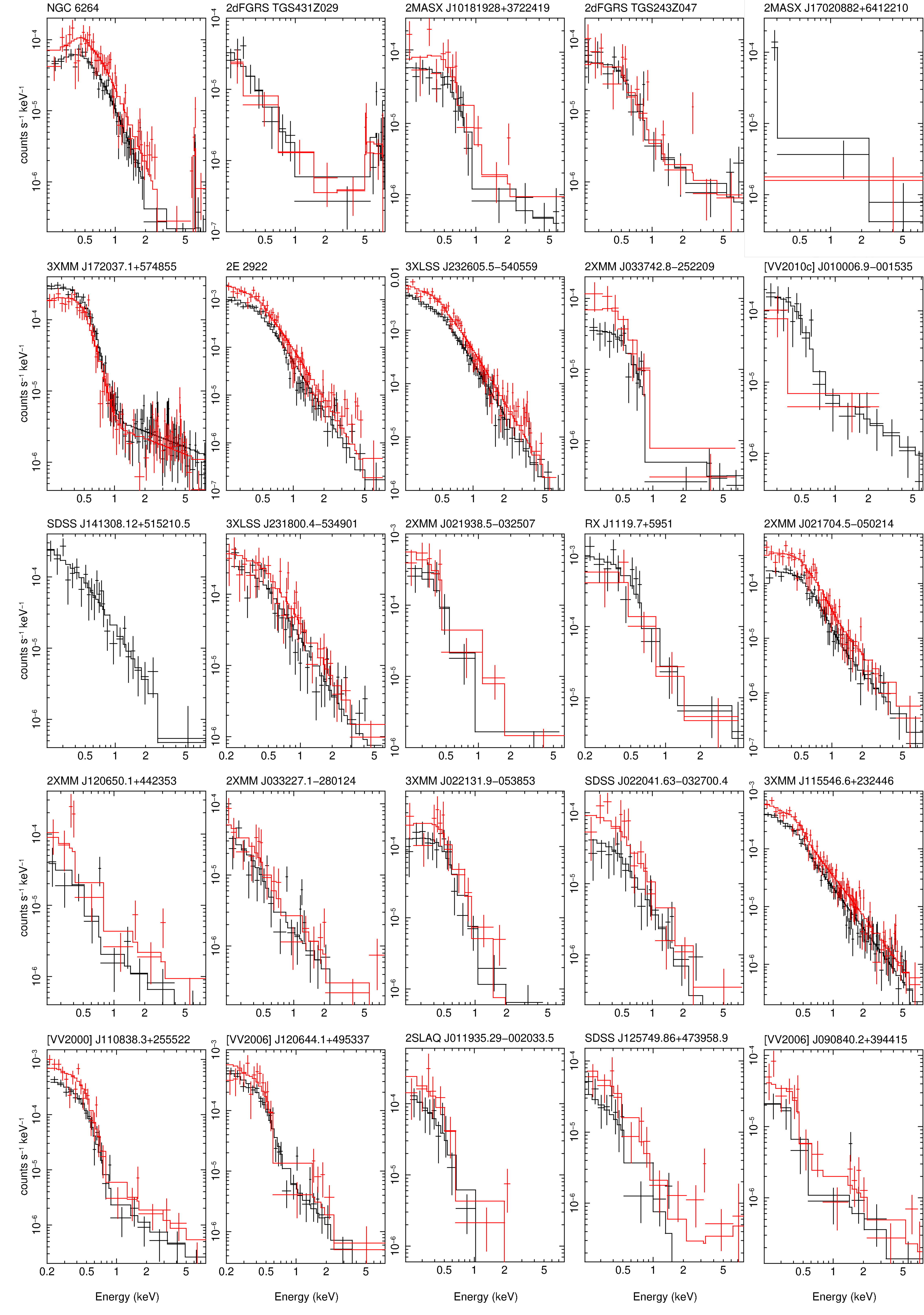}
    \caption{Spectra of the \revv{sources} we found to be "standard" AGN. The colour code is the same as in previous plots. The spectral parameters are listed in Table \ref{tab:non_int}.}
    \label{fig:non_int1}
\end{figure*}

\begin{figure*}
    \centering
	\includegraphics[width=0.9\hsize]{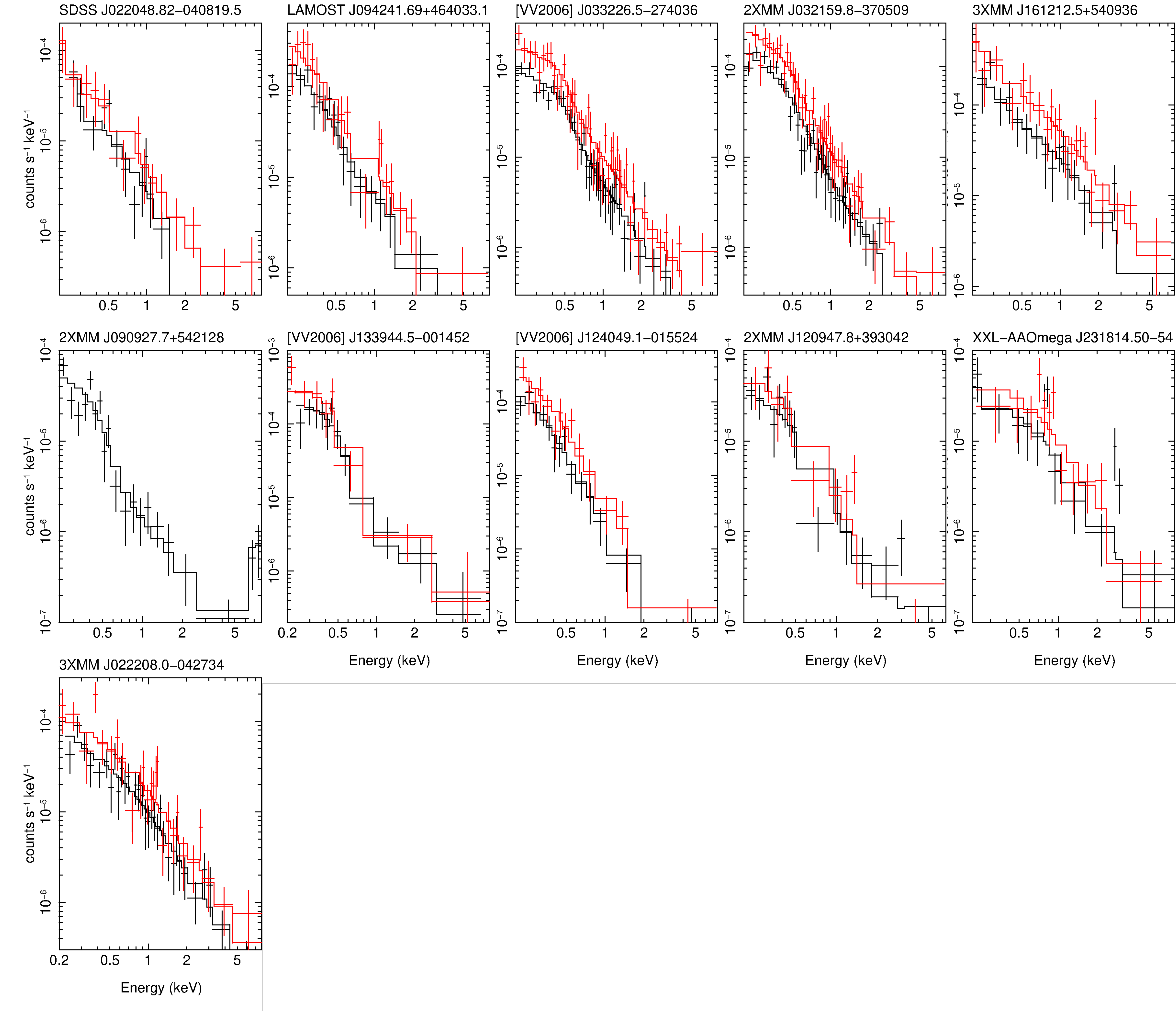}
    \caption{Spectra of the \revv{sources} we found to be "standard" AGN. The colour code is the same as in previous plots. The spectral parameters are listed in Table \ref{tab:non_int}.}
    \label{fig:non_int2}
\end{figure*}

Figures \ref{fig:non_int1} and \ref{fig:non_int2} show the spectra of the sources we did non present in details in the text while Table \ref{tab:non_int} collect the spectral parameters. The model employed is an absorbed blackbody+powerlaw ({\em zTbabs $\times$ (zBbody + zPowerlw)} in XSPEC).
 
\longtab[1]{
\begin{longtable}{|c|c||cccc||r|}
\caption{Spectral parameters of the \revv{sources} we found to be "standard" AGN. Quantities in italic are fixed during the fitting procedure. \rev{The uncertainties reported correspond to a change in the fit statistics of $\Delta C=1$. The unabsorbed luminosity} is computed between 0.5 and 2 keV.}\\
\hline
ID&date&$C/\nu$&$N_\textup{H}$&T&$\Gamma$&$\log_{10}L$\\
(SIMBAD)&&&(10$^{22}$ atoms/cm$^2$)&(keV)&&(erg/s)\\
\hline
\endfirsthead
\caption{Continued.}\\
\hline
ID&date&$C/\nu$&$N_\textup{H}$&T&$\Gamma$&$\log_{10}L$\\
(SIMBAD)&&&(10$^{22}$ atoms/cm$^2$)&(keV)&&(erg/s)\\
\hline
\endhead
\hline
\endfoot
\hline
\endlastfoot
\multirow{3}{*}{NGC  6264}&2005-8-12&9/10&{\it 0.0}&$0.14_{-0.02}^{+0.03}$&-&$41.4_{-0.3}^{+0.2}$\\
&2010-8-26&213/190&{\it 0.0}&$0.14_{-0.01}^{+0.01}$&$2.4_{-0.3}^{+0.3}$&$41.19_{-0.07}^{+0.08}$\\
&2012-2-19&73/81&{\it 0.0}&$0.15_{-0.02}^{+0.02}$&{\it 3.0}&$40.97_{-0.06}^{+0.06}$\\
\hline
\multirow{7}{*}{2dFGRS TGS431Z029}&2003-09-06&17/31&{\it 0.0}&$0.17_{-0.05}^{+0.06}$&{\it -2}&$40.6_{-0.2}^{+0.2}$\\
&2012-05-14&60/79&{\it 0.0}&$0.16_{-0.02}^{+0.02}$&$-1.0_{-1.1}^{+0.9}$&$40.31_{-0.09}^{+0.08}$\\
&2013-11-25&136/118&{\it 0.0}&$0.16_{-0.01}^{+0.01}$&$-2.2_{-0.6}^{+0.5}$&$40.39_{-0.05}^{+0.05}$\\
&2016-05-20&141/129&{\it 0.0}&$0.11_{-0.01}^{+0.02}$&{\it -2.0}&$40.14_{-0.11}^{+0.09}$\\
&2017-05-20&137/121&{\it 0.0}&$0.14_{-0.01}^{+0.01}$&$-1.9_{-0.7}^{+0.6}$&$40.28_{-0.06}^{+0.06}$\\
&2017-11-25&195/129&{\it 0.0}&$0.14_{-0.01}^{+0.01}$&{\it -2.0}&$40.31_{-0.05}^{+0.05}$\\
&2019-11-22&126/124&{\it 0.0}&$0.16_{-0.01}^{+0.01}$&{\it -2.0}&$40.44_{-0.05}^{+0.05}$\\
\hline
2MASX J10181928+3722419&2007-5-18&123/80&{\it 0.0}&$0.11_{-0.01}^{+0.01}$&$0.9_{-0.5}^{+0.5}$&$40.84_{-0.06}^{+0.06}$\\
\hline
\multirow{4}{*}{2dFGRS TGS243Z047}&2018-07-28&103/81&{\it 0.0}&$0.19_{-0.03}^{+0.03}$&$-0.5_{-1.5}^{+0.9}$&$41.15_{-0.08}^{+0.07}$\\
&2018-08-12&150/119&{\it 0.0}&$0.16_{-0.01}^{+0.01}$&$-0.7_{-0.5}^{+0.5}$&$41.21_{-0.04}^{+0.04}$\\
&2018-08-20&105/125&{\it 0.0}&$0.12_{-0.02}^{+0.02}$&$0.8_{-0.5}^{+0.6}$&$41.17_{-0.07}^{+0.06}$\\
&2019-02-09&77/77&{\it 0.0}&$0.15_{-0.01}^{+0.01}$&{\it -2.0}&$41.16_{-0.07}^{+0.07}$\\
\hline
\multirow{6}{*}{2MASX J17020882+6412210}&2000-10-31&21/29&{\it 0.0}&{\it 0.12}&$3.1_{-0.4}^{+0.4}$&$41.0_{-0.2}^{+0.5}$\\
&2002-5-31&76/77&{\it 0.0}&$0.18_{-0.05}^{+0.05}$&-&$41.3_{-0.2}^{+0.1}$\\
&2007-11-14&31/24&{\it 0.0}&$0.4_{-0.1}^{+0.1}$&-&$41.3_{-0.1}^{+0.1}$\\
&2013-5-19&27/26&{\it 0.0}&$0.24_{-0.05}^{+0.06}$&-&$41.2_{-0.1}^{+0.4}$\\
&2013-7-9&81/50&{\it 0.0}&$0.16_{-0.02}^{+0.02}$&$-0.2_{-0.7}^{+0.7}$&$40.82_{-0.08}^{+0.08}$\\
&2013-8-4&30/31&$0.2_{-0.2}^{+0.4}$&$0.11_{-0.04}^{+0.06}$&-&$41.5_{-0.6}^{+0.4}$\\
\hline
\multirow{4}{*}{3XMM J172037.1+574855}&2009-8-28&147/122&{\it 0.0}&$0.13_{-0.01}^{+0.01}$&$0.6_{-0.3}^{+0.3}$&$41.95_{-0.04}^{+0.04}$\\
&2015-3-28&103/81&{\it 0.0}&$0.12_{-0.01}^{+0.01}$&$-0.1_{-0.3}^{+0.3}$&$42.08_{-0.03}^{+0.03}$\\
&2015-6-15&326/207&$0.02_{-0.02}^{+0.02}$&$0.09_{-0.0}^{+0.0}$&$0.5_{-0.1}^{+0.1}$&$42.45_{-0.04}^{+0.05}$\\
&2015-9-3&210/128&{\it 0.0}&$0.10_{-0.004}^{+0.005}$&$0.6_{-0.1}^{+0.1}$&$42.48_{-0.01}^{+0.01}$\\
\hline
2E 2922&2011-1-22&240/165&{\it 0.0}&$0.113_{-0.003}^{+0.004}$&$2.1_{-0.2}^{+0.2}$&$43.12_{-0.01}^{+0.01}$\\
\hline
\multirow{2}{*}{3XLSS J232605.5-540559}&2008-4-17&164/169&{\it 0.0}&$0.13_{-0.01}^{+0.01}$&$2.8_{-0.1}^{+0.1}$&$43.8_{-0.01}^{+0.01}$\\
&2012-11-25&106/89&{\it 0.0}&$0.11_{-0.01}^{+0.01}$&$1.9_{-0.4}^{+0.5}$&$42.93_{-0.04}^{+0.03}$\\
\hline
2XMM J033742.8-252209&2001-8-18&108/117&{\it 0.0}&$0.14_{-0.01}^{+0.01}$&$-0.1_{-0.4}^{+0.4}$&$42.33_{-0.05}^{+0.05}$\\
\hline
[VV2010c] J010006.9-001535&2014-7-4&26/36&{\it 0.0}&$0.09_{-0.01}^{+0.01}$&$1.1_{-0.3}^{+0.3}$&$42.69_{-0.06}^{+0.06}$\\
\hline
SDSS J141308.12+515210.5&2017-5-11&56/63&{\it 0.0}&{\it 0.12}&$2.3_{-0.7}^{+0.5}$&$43.16_{-0.04}^{+0.04}$\\
\hline
\multirow{2}{*}{3XLSS J231800.4-534901}&2009-12-5&83/62&{\it 0.0}&$0.19_{-0.02}^{+0.02}$&$1.4_{-0.8}^{+1.2}$&$43.45_{-0.05}^{+0.05}$\\
&2012-10-27&159/119&{\it 0.0}&$0.10_{-0.02}^{+0.02}$&$2.2_{-0.2}^{+0.2}$&$43.36_{-0.03}^{+0.03}$\\
\hline
\multirow{2}{*}{2XMM J021938.5-032507}&2007-1-11&28/38&$0.5_{-0.3}^{+0.5}$&$0.05_{-0.01}^{+0.01}$&$2.8_{-0.7}^{+1.1}$&$45.0_{-0.9}^{+1.6}$\\
&2009-1-3&31/31&$0.9_{-0.3}^{+0.9}$&$0.05_{-0.01}^{0.02}$&$2.8_{-1.3}^{+2.2}$&$45.3_{-1.3}^{+2.5}$\\
\hline
\multirow{2}{*}{RX J1119.7+5951}&2007-5-20&37/35&{\it 0.0}&$0.13_{-0.02}^{+0.02}$&$1.5_{-0.9}^{+1.0}$&$44.26_{-0.06}^{+0.06}$\\
&2017-8-12&38/33&{\it 0.0}&$0.012_{-0.004}^{+0.003}$&$2.7_{-0.2}^{+0.2}$&$43.94_{-0.07}^{+0.07}$\\
\hline
\multirow{5}{*}{2XMM J021704.5-050214}&2000-7-31&95/109&{\it 0.0}&$0.20_{-0.03}^{+0.04}$&$3.1_{-1.6}^{+3.2}$&$43.19_{-0.06}^{+0.05}$\\
&2000-8-2&83/114&{\it 0.0}&{\it 0.12}&$2.4_{-0.3}^{+0.4}$&$43.22_{-0.05}^{+0.04}$\\
&2002-8-12&197/157&{\it 0.0}&$0.17_{-0.01}^{+0.01}$&$2.2_{-0.2}^{+0.2}$&$43.81_{-0.02}^{+0.02}$\\
&2015-7-5&22/31&{\it 0.0}&{\it 0.12}&$2.1_{-0.4}^{+0.5}$&$43.0_{-0.2}^{+0.2}$\\
&2015-9-18&41/57&{\it 0.0}&$0.43_{-0.06}^{+0.09}$&$1.7_{-1.7}^{+0.9}$&$42.84_{-0.09}^{+0.12}$\\
\hline
2XMM J120650.1+442353&2003-6-11&71/70&{\it 0.0}&$0.11_{-0.02}^{+0.02}$&$1.0_{-0.3}^{+0.3}$&$42.9_{-0.1}^{+0.1}$\\
\hline
\multirow{11}{*}{2XMM J033227.1-280124}&2001-7-27&109/92&{\it 0.0}&$0.08_{-0.02}^{+0.02}$&$1.4_{-0.3}^{+0.3}$&$42.99_{-0.08}^{+0.07}$\\
&2002-1-13&103/96&{\it 0.0}&$0.1_{-0.02}^{+0.02}$&$2.4_{-0.4}^{+0.4}$&$43.26_{-0.04}^{+0.04}$\\
&2002-1-14&78/90&{\it 0.0}&$0.08_{-0.01}^{+0.01}$&$0.8_{-0.6}^{+0.7}$&$42.84_{-0.06}^{+0.05}$\\
&2002-1-16&131/129&{\it 0.0}&$0.11_{-0.01}^{+0.01}$&$1.7_{-0.3}^{+0.3}$&$43.38_{-0.03}^{+0.03}$\\
&2002-1-17&16/36&{\it 0.0}&$0.06_{-0.01}^{+0.02}$&$1.3_{-0.6}^{+0.6}$&$42.9_{-0.2}^{+0.1}$\\
&2002-1-23&110/85&{\it 0.0}&$0.22_{-0.02}^{+0.02}$&-&$42.94_{-0.08}^{+0.07}$\\
&2004-11-17&257/173&{\it 0.0}&$0.59_{-0.05}^{+0.04}$&-&$43.37_{-0.05}^{+0.05}$\\
&2009-1-12&103/89&{\it 0.0}&$0.11_{-0.02}^{+0.02}$&$2.0_{-0.3}^{+0.4}$&$42.98_{-0.03}^{+0.03}$\\
&2010-1-26&139/154&{\it 0.0}&$0.09_{-0.02}^{+0.02}$&$1.7_{-0.3}^{+0.3}$&$42.82_{-0.05}^{+0.04}$\\
&2018-8-12&95/80&{\it 0.0}&$0.04_{-0.02}^{+0.02}$&$1.9_{-0.3}^{+0.3}$&$42.91_{-0.08}^{+0.07}$\\
&2019-8-6&39/37&{\it 0.0}&$0.12_{-0.03}^{+0.04}$&$0.3_{-2.7}^{+1.0}$&$42.7_{-0.1}^{+0.1}$\\
\hline
\multirow{5}{*}{3XMM J022131.9-053853}&2003-7-24&5/5&{\it 0.0}&$0.08_{-0.02}^{+0.02}$&$0.4_{-2.1}^{+1.6}$&$44.2_{-0.1}^{+0.1}$\\
&2006-7-6&51/50&{\it 0.0}&$0.11_{-0.01}^{+0.01}$&$0.9_{-0.5}^{+0.4}$&$44.03_{-0.06}^{+0.06}$\\
&2006-7-14&35/33&{\it 0.0}&$0.09_{-0.01}^{+0.02}$&$0.6_{-0.5}^{+0.6}$&$43.8_{-0.2}^{+0.2}$\\
&2009-1-1&86/55&{\it 0.0}&$0.15_{-0.01}^{+0.01}$&$2.2_{-0.6}^{+0.6}$&$44.02_{-0.05}^{+0.05}$\\
&2016-1-8&8/13&{\it 0.0}&$0.15_{-0.03}^{+0.03}$&-&$43.88_{-0.09}^{+0.08}$\\
\hline
SDSS J022041.63-032700.4&2002-8-15&68/74&$0.13_{-0.10}^{+0.31}$&$0.15_{-0.05}^{+0.04}$&$2.1_{-0.5}^{+0.6}$&$43.47_{-0.06}^{+0.05}$\\
\hline
\multirow{8}{*}{3XMM J115546.6+232446}&2000-6-23&4/4&{\it 0.0}&{\it 0.12}&-&$44.1_{-0.2}^{+0.2}$\\
&2000-12-6&33/31&{\it 0.0}&{\it 0.12}&$2.1_{-0.6}^{+0.7}$&$43.77_{-0.10}^{+0.08}$\\
&2004-3-6&134/109&{\it 0.0}&$0.12_{-0.04}^{+0.04}$&$1.9_{-0.2}^{+0.2}$&$43.46_{-0.06}^{+0.06}$\\
&2007-7-10&1/1&{\it 0.0}&{\it 0.12}&-&$43.9_{-0.4}^{+0.3}$\\
&2007-11-26&75/76&{\it 0.0}&$0.11_{-0.01}^{+0.01}$&$2.0_{-0.3}^{+0.3}$&$43.8_{-0.02}^{+0.02}$\\
&2007-12-12&174/151&{\it 0.0}&$0.11_{-0.01}^{+0.01}$&$2.2_{-0.2}^{+0.3}$&$43.65_{-0.02}^{+0.02}$\\
&2008-12-5&256/214&{\it 0.0}&$0.1_{-0.01}^{+0.01}$&$2.4_{-0.1}^{+0.1}$&$44.31_{-0.01}^{+0.01}$\\
&2008-12-7&232/193&{\it 0.0}&$0.1_{-0.01}^{+0.01}$&$2.3_{-0.1}^{+0.1}$&$44.03_{-0.02}^{+0.02}$\\
\hline
[VV2000] J110838.3+255522&2017-5-30&135/113&{\it 0.0}&$0.124_{-0.004}^{+0.003}$&$1.0_{-0.3}^{+0.3}$&$44.29_{-0.02}^{+0.02}$\\
\hline
[VV2006] J120644.1+495337&2017-11-21&122/96&$0.5_{-0.2}^{+0.2}$&$0.08_{-0.01}^{+0.02}$&$2.3_{-0.3}^{+0.3}$&$44.56_{-0.02}^{+0.02}$\\
\hline
2SLAQ J011935.29-002033.5&2014-7-20&22/33&{\it 0.0}&$0.12_{-0.01}^{+0.02}$&$1.9_{-0.7}^{+0.7}$&$44.12_{-0.08}^{+0.07}$\\
\hline
\multirow{3}{*}{SDSS J125749.86+473958.9}&2007-5-24&178/91&{\it 0.0}&$0.11_{-0.02}^{+0.02}$&$1.9_{-0.4}^{+0.5}$&$43.49_{-0.07}^{+0.07}$\\
&2007-5-26&45/51&{\it 0.0}&$0.14_{-0.01}^{+0.01}$&$1.9_{-1.5}^{+2.6}$&$43.72_{-0.09}^{+0.08}$\\
&2008-12-18&45/43&{\it 0.0}&$0.11_{-0.02}^{+0.03}$&$1.6_{-0.8}^{+0.8}$&$43.2_{-0.2}^{+0.1}$\\
\hline
[VV2006] J090840.2+394415&2006-10-31&141/101&{\it 0.0}&$0.11_{-0.01}^{+0.02}$&$1.1_{-0.3}^{+0.3}$&$43.3_{-0.1}^{+0.1}$\\
\hline
\multirow{3}{*}{SDSS J022048.82-040819.5}&2003-1-25&20/26&{\it 0.0}&{\it 0.12}&$3.2_{-0.9}^{+1.2}$&$44.1_{-0.3}^{+0.2}$\\
&2016-7-5&54/72&{\it 0.0}&$0.03_{-0.01}^{+0.11}$&$2.6_{-0.1}^{+0.2}$&$43.77_{-0.07}^{+0.07}$\\
&2016-7-7&27/29&{\it 0.0}&$0.18_{-0.05}^{+0.07}$&-&$43.8_{-0.1}^{+0.1}$\\
\hline
LAMOST J094241.69+464033.1&2010-4-17&148/150&{\it 0.0}&$0.11_{-0.02}^{+0.02}$&$2.1_{-0.4}^{+0.4}$&$44.28_{-0.05}^{+0.05}$\\
\hline
\multirow{4}{*}{[VV2006] J033226.5-274036}&2001-7-27&262/155&{\it 0.0}&$0.06_{-0.02}^{+0.02}$&$2.7_{-0.1}^{+0.1}$&$44.57_{-0.02}^{+0.02}$\\
&2002-1-13&204/149&{\it 0.0}&$0.13_{-0.01}^{+0.01}$&$2.2_{-0.1}^{+0.1}$&$44.55_{-0.02}^{+0.02}$\\
&2002-1-23&357/166&{\it 0.0}&$0.11_{-0.02}^{+0.01}$&$2.5_{-0.1}^{+0.1}$&$44.61_{-0.01}^{+0.01}$\\
&2010-1-26&232/199&{\it 0.0}&$0.16_{-0.01}^{+0.01}$&$2.0_{-0.2}^{+0.2}$&$44.23_{-0.02}^{+0.02}$\\
\hline
\multirow{5}{*}{2XMM J032159.8-370509}&2001-4-17&112/95&{\it 0.0}&$0.58_{-0.07}^{+0.16}$&-&$44.44_{-0.08}^{+0.08}$\\
&2005-8-11&138/191&$0.5_{-0.2}^{+0.2}$&$0.08_{-0.01}^{+0.02}$&$2.8_{-0.3}^{+0.3}$&$45.6_{-0.6}^{+0.5}$\\
&2007-8-19&128/141&{\it 0.0}&$0.10_{-0.01}^{+0.01}$&$1.5_{-0.3}^{+0.3}$&$44.13_{-0.06}^{+0.06}$\\
&2009-6-25&139/152&{\it 0.0}&$0.15_{-0.06}^{+0.11}$&$1.2_{-0.5}^{+0.5}$&$42.8_{-0.2}^{+0.2}$\\
&2019-4-20&15/15&{\it 0.0}&-&$1.3_{-1.0}^{+1.1}$&$43.6_{-0.6}^{+0.6}$\\
\hline
\multirow{5}{*}{3XMM J161212.5+540936}&2002-9-12&109/79&{\it 0.0}&$0.23_{-0.04}^{+0.03}$&$1.3_{-0.5}^{+0.5}$&$44.73_{-0.05}^{+0.05}$\\
&2002-9-14&93/72&{\it 0.0}&$0.14_{-0.04}^{+0.04}$&$1.9_{-0.3}^{+0.3}$&$44.69_{-0.06}^{+0.06}$\\
&2002-9-18&92/110&{\it 0.0}&$0.07_{-0.04}^{+0.05}$&$1.8_{-0.2}^{+0.1}$&$44.53_{-0.06}^{+0.06}$\\
&2005-12-7&32/26&{\it 0.0}&$1.1_{-0.1}^{+0.2}$&-&$44.5_{-0.2}^{+0.2}$\\
&2006-3-8&44/38&{\it 0.0}&$0.7_{-0.1}^{+0.2}$&-&$44.8_{-0.1}^{+0.1}$\\
\hline
\multirow{2}{*}{2XMM J090927.7+542128}&2001-4-29&137/111&{\it 0.0}&$0.13_{-0.06}^{+0.07}$&$2.0_{-0.2}^{+0.2}$&$44.05_{-0.08}^{+0.08}$\\
&2005-3-28&56/55&{\it 0.0}&$0.13_{-0.01}^{+0.01}$&$1.9_{-0.4}^{+0.4}$&$44.15_{-0.06}^{+0.06}$\\
\hline
\multirow{2}{*}{[VV2006] J133944.5-001452}&2005-1-12&28/42&{\it 0.0}&$0.05_{-0.02}^{+0.03}$&$2.4_{-0.6}^{+0.5}$&$45.1_{-0.2}^{+0.2}$\\
&2005-7-17&62/60&{\it 0.0}&$0.17_{-0.01}^{+0.01}$&$1.2_{-0.2}^{+0.3}$&$44.77_{-0.07}^{+0.07}$\\
\hline
\multirow{2}{*}{[VV2006] J124049.1-015524}&2006-1-7&65/74&{\it 0.0}&$0.09_{-0.03}^{+0.03}$&$3.3_{-0.6}^{+0.5}$&$44.56_{-0.05}^{+0.05}$\\
&2017-12-14&43/54&{\it 0.0}&{\it 0.12}&$2.1_{-0.7}^{+0.7}$&$43.9_{-0.2}^{+0.2}$\\
\hline
2XMM J120947.8+393042&2000-12-22&102/73&{\it 0.0}&$0.12_{-0.01}^{+0.01}$&$1.7_{-0.3}^{+0.3}$&$44.2_{-0.1}^{+0.1}$\\
\hline
\multirow{3}{*}{XXL-AAOmega J231814.50-544112.3}&2007-10-27&74/86&{\it 0.0}&$0.39_{-0.06}^{+0.07}$&$1.4_{-0.6}^{+0.5}$&$44.18_{-0.09}^{+0.08}$\\
&2009-12-5&26/37&{\it 0.0}&$0.4_{-0.1}^{+0.1}$&-&$44.2_{-0.3}^{+0.2}$\\
&2012-12-7&42/49&{\it 0.0}&$0.8_{-0.1}^{+0.1}$&-&$43.8_{-0.1}^{+0.1}$\\
\hline
\multirow{4}{*}{3XMM J022208.0-042734}&2002-1-31&201/95&{\it 0.0}&$0.85_{-0.07}^{+0.07}$&-&$44.49_{-0.09}^{+0.09}$\\
&2002-8-14&26/37&{\it 0.0}&{\it 0.12}&$2.2_{-0.2}^{+0.2}$&$45.0_{-0.1}^{+0.1}$\\
&2016-7-7&63/89&{\it 0.0}&$0.15_{-0.05}^{+0.06}$&$1.7_{-0.2}^{+0.2}$&$44.6_{-0.07}^{+0.07}$\\
&2016-7-29&137/119&$0.19_{-0.19}^{+0.44}$&$0.08_{-0.03}^{+0.04}$&$2.2_{-0.2}^{+0.3}$&$44.74_{-0.04}^{+0.04}$\\
\hline
\end{longtable}
\label{tab:non_int}
}

\end{document}